\documentclass[12pt]{article}
\usepackage[latin1]{inputenc}
\usepackage[british]{babel}
\usepackage{cmap}
\usepackage{lmodern}

\usepackage{amssymb, amsmath, amsthm}
\usepackage[a4paper,top=25mm,bottom=25mm,left=25mm,right=25mm]{geometry}
\usepackage{etex}
\usepackage{ragged2e}

\usepackage{authblk} 
\usepackage{pifont}
\usepackage{graphicx}
\usepackage[usenames,dvipsnames,svgnames,table]{xcolor}
\usepackage[figuresright]{rotating}
\usepackage{xtab} 
\usepackage{longtable} 
\usepackage{multirow}
\usepackage{footnote}
\usepackage[stable]{footmisc}
\usepackage{chngpage} 
\usepackage{pdflscape} 
\usepackage[nottoc,notlot,notlof]{tocbibind} 

\usepackage{pgfplots}
\pgfplotsset{compat=1.14}
\usepackage{setspace}

\makesavenoteenv{tabular}
\usepackage{tabularx}
\usepackage{booktabs}
\usepackage{threeparttable}
\usepackage[referable]{threeparttablex} 
\newcolumntype{R}{>{\raggedleft\arraybackslash}X}
\newcolumntype{L}{>{\raggedright\arraybackslash}X}
\newcolumntype{C}{>{\centering\arraybackslash}X}
\newcolumntype{A}{>{\columncolor{gray!25}}C}
\newcolumntype{a}{>{\columncolor{gray!25}}c}

\newlength{\tablen}

\usepackage{dcolumn} 
\newcolumntype{.}{D{.}{.}{-1}}

\usepackage{tikz}
\usetikzlibrary{arrows, calc, matrix, patterns, positioning, trees}
\usepackage[semicolon]{natbib}
\usepackage[hyphens]{url}
\usepackage{hyperref} 
\hypersetup{
  colorlinks   = true,    
  urlcolor     = blue,    
  linkcolor    = blue,    
  citecolor    = red      
}
\usepackage{microtype}
\usepackage[justification=centering]{caption} 

\usepackage[labelformat=simple]{subcaption}

\DeclareCaptionLabelFormat{parenthesis}{(#2)}
\makeatletter
\renewcommand\p@subfigure{\arabic{figure}.}
\makeatother

\DeclareCaptionLabelFormat{parenthesis}{(#2)}
\makeatletter
\renewcommand\p@subtable{\arabic{table}.}
\makeatother

\usepackage{enumitem}

\setlist[itemize]{leftmargin=2.5\parindent}
\setlist[enumerate]{leftmargin=2.5\parindent}

\newcommand\Pair[3]{%
  \begin{tabular}{|>{\centering\arraybackslash}m{0.75cm}|>{\centering\arraybackslash}m{1.5cm}|}
  \hline
  \multirow{2}{*}{#1} & #2 \\ \cline{2-2}
   & #3 \\
  \hline
  \end{tabular}%
}

\theoremstyle{plain}

\theoremstyle{definition}

\theoremstyle{remark}

\def\keywords{\vspace{.5em} 
{\noindent \textit{Keywords}: }}

\def\JEL{\vspace{.5em} 
{\noindent \textbf{\emph{JEL} classification number}: }}

\def\AMS{\vspace{.5em} 
{\noindent \textbf{\emph{MSC} class}: }}

\author{\href{https://sites.google.com/site/laszlocsato87}{L\'aszl\'o Csat\'o}\thanks{~E-mail: \emph{laszlo.csato@uni-corvinus.hu}} }
\affil{Institute for Computer Science and Control, Hungarian Academy of Sciences (MTA SZTAKI) \\
Laboratory on Engineering and Management Intelligence, Research Group of Operations Research and Decision Systems}
\affil{Corvinus University of Budapest (BCE) \\
Department of Operations Research and Actuarial Sciences}
\affil{Budapest, Hungary}
\title{A simulation comparison of tournament designs for the World Men's Handball Championships}
\date{\today}

\def\Dedication{
{\noindent
$\mathfrak{Wo}$ $\mathfrak{es}$ $\mathfrak{auch}$ $\mathfrak{kein}$ $\mathfrak{System,}$ $\mathfrak{keinen}$ $\mathfrak{Wahrheitsapparat}$ $\mathfrak{gibt}$, $\mathfrak{da}$ $\mathfrak{gibt}$ $\mathfrak{es}$ $\mathfrak{doch}$ $\mathfrak{eine}$ $\mathfrak{Wahrheit}$, $\mathfrak{und}$ $\mathfrak{diese}$ $\mathfrak{wird}$ $\mathfrak{dann}$ $\mathfrak{meistens}$ $\mathfrak{nur}$ $\mathfrak{durch}$ $\mathfrak{ein}$ $\mathfrak{ge\ddot{u}btes}$ $\mathfrak{Urteil}$ $\mathfrak{und}$ $\mathfrak{den}$ $\mathfrak{Takt}$ $\mathfrak{einer}$ $\mathfrak{langen}$ $\mathfrak{Erfahrung}$ $\mathfrak{gefunden}$. $\mathfrak{Gibt}$ $\mathfrak{also}$ $\mathfrak{die}$ $\mathfrak{Geschichte}$ $\mathfrak{hier}$ $\mathfrak{keine}$ $\mathfrak{Formeln}$, $\mathfrak{so}$ $\mathfrak{gibt}$ $\mathfrak{sie}$ $\mathfrak{doch}$ $\mathfrak{hier}$ $\mathfrak{wie}$ $\mathfrak{\ddot{u}berall}$ $\mathfrak{\ddot{U}bung}$ $\mathfrak{des}$ $\mathfrak{Urteils}$.\footnote{~
``Where neither system nor any dogmatic apparatus can be found, there may still be truth, and this truth will then, in most cases, only be discovered by a practised judgment and the tact of long experience. Therefore, even if history does not here furnish any formula, we may be certain that here as well as everywhere else, it will give us \emph{exercise for the judgment}.'' (Source: Carl von Clausewitz: \emph{On War}, Book 6, Chapter 30 -- Defence of a theatre of war (continued): When no decision is sought for. Translated by Colonel James John Graham, London, N. Tr\"ubner, 1873. \url{http://clausewitz.com/readings/OnWar1873/TOC.htm})}
}
\vspace{0.25cm}

\flushright
\noindent (Carl von Clausewitz: \emph{Vom Kriege})

\vspace{1cm} 
\justify }

\begin{document}

\maketitle

\Dedication

\begin{abstract}
\noindent
The study aims to compare different designs for the World Men's Handball Championships. This event, organised in every two years, has adopted four hybrid formats consisting of knockout and round-robin stages in recent decades, including a change of design between the two recent championships in 2017 and 2019. They are evaluated under two extremal seeding policies with respect to various outcome measures through Monte-Carlo simulations.
We find that the ability to give the first four positions to the strongest teams, as well as the expected quality and outcome uncertainty of the final is not necessarily a monotonic function of the number of matches played: the most frugal format is the second best with respect to these outcome measures, making it a good compromise in an unavoidable trade-off. A possible error is identified in a particular design.
The relative performance of the formats is independent of the seeding rules and the competitive balance of the teams. The recent reform is demonstrated to have increased the probability of winning for the top teams. Our results have useful implications for the organisers of hybrid tournaments.

\JEL{C44, C63, Z20}

\AMS{62F07, 68U20}

\end{abstract}

\vspace{4cm}
\tableofcontents

\clearpage
\hspace{0pt}
\vfill

\listoftables

\vspace{4cm}
\listoffigures

\vfill
\hspace{0pt}
\clearpage

\section{Introduction} \label{Sec1}

Finding the optimal design of sports tournaments is an important question of scientific research \citep{Szymanski2003}.
Ignoring the assignment of referees \citep{AlarconDuranGuajardo2014, AtanHuseyinoglu2017} -- who may be biased, for example, towards the home team \citep{GaricanoPalacios-HuertaPrendergast2005} --, organisers and sports governing bodies have supposedly no influence on match outcomes.
However, they can certainly choose other characteristics of a tournament, including the format \citep{ScarfYusofBilbao2009, ScarfYusof2011, Guyon2018a}, the schedule of individual matches \citep{Ribeiro2012, AtanHuseyinoglu2017, DuranGuajardoSaure2017}, the seeding policy \citep{Guyon2015, LalienaLopez2018}, or the progression rules \citep{Csato2018h, Csato2018b, Csato2019i, Csato2019b, DagaevSonin2018, Vong2017}.

The current paper considers tournament designs as probabilistic mechanisms that select high-quality alternatives (players or teams) in a noisy environment \citep{Ryvkin2010}.
Operational Research (OR) can play a prominent role here by analysing the effects of different competition structures on particular aspects of the sporting event: given a particular metric as an objective, and respecting design constraints, it becomes possible to choose the most favourable version.

In sports involving pairwise matches, there are two fundamental tournament formats \citep{ScarfYusofBilbao2009}. The first is the \emph{knockout} tournament where matches are played in rounds such that the winners play against each other in the next round, while the losers are immediately eliminated from the tournament. The sole remaining player, the winner of the final gets the first prize.
The second basic design is the \emph{round-robin} tournament where every competitor plays every other such that they earn points based on their number of wins, draws, and losses. The winner is the team with the greatest point score.
All other designs can be considered as variations, such as the \emph{double elimination} \citep{McGarrySchutz1997, StantonWilliams2013}, the \emph{Swiss system} \citep{Appleton1995, Csato2013a, Csato2017c}, and hybrids like the FIFA World Cup or the UEFA Champions League in association football.

Tournament success measures can be defined in a relatively straightforward way.
On the other hand, the identification of design constraints is usually more complicated because they are rarely communicated by the administrators. The only plausible assumption seems to be that a format used in the past for a given tournament remains feasible in the future.
However, this consideration does not help much when the tournament receives a modification to its structure only in parallel with a change in the number of competitors. For example, FIFA World Cup was expanded to 24 teams in 1982, then to 32 in 1998, while the 2026 World Cup will have 48 finalist teams, but its format has remained the same for the same number of teams in these years. Similarly, the biannual European Men's and Women's Handball Championships started with 12 teams in 1994, and were expanded to 16 teams in 2002, but were organised according to the same structure for a given number of competitors. It means that suggesting a novel design has not much practical value unless it dominates the one applied in the real-world in (almost) every respect.

In contrast, some high profile events have received a regular modification to their structure. We will analyse here a probably unique example, the \href{https://en.wikipedia.org/wiki/IHF_World_Men\%27s_Handball_Championship}{IHF World Men's Handball Championship}. This event is held in every two years since 1993 and is one of the most important and prestigious championships for men's handball national teams along Olympic games and the \href{https://en.wikipedia.org/wiki/European_Men\%27s_Handball_Championship}{EHF European Men's Handball Championship} as handball is most popular in the countries of continental Europe, which have won all medals but one in the World Men's Championships. Attendance of the \href{https://en.wikipedia.org/wiki/2019_World_Men\%27s_Handball_Championship}{Championship in 2019}, hosted by Denmark and Germany, was over $900{,}000$, more than $9{,}000$ per match.

The number of qualified teams has remained fixed at 24 since 1995, but the tournament format has changed several times over the last two decades. Between 1995 and 2001 (four events), there were group games in the preliminary round, followed by a knockout stage. This format was used again between 2013 and 2017 (three events). However, there were two subsequent group stages between 2003 and 2011 (five events), in three different variants, one of them returning in 2019.
To conclude, there are four tournament structures implemented in recent years, including a change between the two recent tournaments. This indicates that the organisers experiment with finding the best design, which offers an extraordinary opportunity to compare them with the tools of OR.

It is clear that such complex designs, consisting of knock-out and round-robin stages, can be analysed only via Monte-Carlo simulations. Academic literature has made several attempts to address similar problems.
\citet{ScarfYusofBilbao2009} propose a number of tournament metrics and describe how they may be evaluated for a particular design. The authors use the UEFA Champions League to illustrate their methodology. \citet{ScarfYusof2011} extend this investigation by considering the effect of the seeding policy on outcome uncertainty while taking competitive balance into account. 
\citet{GoossensBelienSpieksma2012} examine four league formats that have been considered by the Royal Belgian Football Association. 
\citet{LasekGagolewski2015} compare the recently introduced competition format for the top association football division in Poland to the standard double round-robin structure.
\citet{Yusofetal2016} develop a system called `E-compare of Soccer Tournament Structures' to assist decision makers in determining the fairest design for association football tournaments.
\citet{LasekGagolewski2018} examine the efficacy of league formats in ranking football teams and find that the performance of formats consisting of round-robin stages mainly depends on the total number of matches played.
\citet{DagaevRudyak2019} assess a recent reform of the seeding system in the UEFA Champions League.
\citet{Csato2019h} evaluates an alternative of the traditional multi-stage tournament design through the example of the EHF Champions League, the most prestigious men's handball club competition in Europe.

Most of these papers use specific models for simulating match results, however, we want to avoid the use of such sophisticated assumptions to compare the tournament formats for a number of reasons.
First, we follow general works on the efficacy of sports tournaments \citep{Appleton1995, McGarrySchutz1997} or ranking methods \citep{MendoncaRaghavachari2000}, which apply this choice.
Second, at least according to our knowledge, there exists no particular prediction model fitted to handball results, contrary to the variety of methods making a good prediction on the outcome of a single match between two football teams \citep{Maher1982, DixonColes1997, KoningKoolhaasRenesRidder2003}. The main difficulty is probably that handball is a fast, dynamic and high-scoring game, where professional teams now typically score between 20 and 35 goals each, therefore the technical analysis of a handball match poses a serious challenge \citep{Bilge2012, GruicVuletaMilanovic2007}. According to \citet{DumanganeRosatiVolossovitch2009}, the dynamics of handball matches violate both independence and identical distribution, in some cases having a non-stationary behaviour. In addition, some tournament designs analysed here have been applied only once, so the lack of historical data prevents fitting a specific prediction model.
Third, \citet{KrumerMegidishSela2017a} prove that in round-robin tournaments among three or four symmetric contestants, there is a first-mover advantage driven by strategic effects arising from the subgame perfect equilibrium, while \citet{KrumerLechner2017} give an empirical proof of this finding. Since all of our designs contain at least one group stage, even the schedule of the matches may influence the outcome of the tournament.

To summarise, the exact modelling of handball matches organised in such complicated hybrid designs seems to be beyond the current knowledge of the academic community.
However, since our intention is only to compare the tournament formats, and not to estimate the chance of winning, a number of models within reason could be taken to determine the winners \citep{Appleton1995}. Nonetheless, this implies that all calculations are for comparative purposes only.

The main contribution and novelty of our research is the analysis of a particular -- but by no means marginal -- handball tournament by simulations, which has received several modifications to its format recently, indicating that the organisers are probably uncertain on its appropriate design. While the choice of tournament format is driven by a number of factors \citep{Szymanski2003, Wright2014}, we focus on its ability to give the first four positions to the strongest teams, and on the quality and competitive balance of the championship final. Consequently, in the following, a design will be called more \emph{efficacious} if it performs better with respect to \emph{all} these criteria -- which can be conflicting, hence this definition of efficacy is not guaranteed to produce a strict ranking of all formats.

Note that our definition does not coincide with the standard meaning of efficacy, the ability of a tournament to produce accurate rankings with respect to teams' true abilities. However, in our view, the latter approach is reasonable only in round-robin tournaments \citep{LasekGagolewski2018}, or if the number of competitors is small \citep{McGarrySchutz1997, MendoncaRaghavachari2000}. Now the ranking outside the top four is unreliable and almost irrelevant as all designs are centred around the semifinals, while the tournament final has a prominent role in creating media attention, so taking only the ranking ability into account is not enough if the key determinants of demand \citep{BorlandMacDonald2003} are unfavourable for the most important match.

We have some surprising findings, for example, the most frugal design in the number of matches played is the second best with respect to efficacy, thus it seems to be a good compromise in the unavoidable trade-off. This is mainly caused by the smaller groups of four teams each instead of six in the first round-robin stage, a suggestion is worth further consideration. Our calculations also reveal that the recent format change of the World Men's Handball Championship has increased the probability of winning for the top teams.

In short, the results will have useful implications for hybrid tournaments that are applied in several sports such as basketball, handball, and volleyball, some of them are presented at the end of the paper.

The paper is structured as follows.
Section~\ref{Sec2} describes the tournament designs, the metrics used for the comparison of different formats, and the simulation experiment.
The results and their sensitivity analysis are detailed in Section~\ref{Sec3}.
Finally, Section~\ref{Sec4} discusses our main findings and concludes.

\section{Methodology} \label{Sec2}

For the comparison of different tournament designs, it is necessary to use simulations as historical data are limited because some formats were applied only once.

\subsection{Tournament designs} \label{Sec21}

The \href{https://en.wikipedia.org/wiki/IHF_World_Men\%27s_Handball_Championship}{IHF World Men's Handball Championships} have been organised with $24$ participating teams in four fundamentally different designs in recent decades. Our investigation is restricted to these tournament formats in order to avoid the question of whether the suggested design can be implemented in practice.

Each format contains one or two group stages. Groups are round-robin tournaments with all teams playing once against any other team in their group. In the case of two group stages, the results of the matches played in the preliminary round between teams of the same main round group are carried over to the main round \citep{Csato2019i}.

Organisers provide a strict final ranking at the end of the tournament, meaning that usually there are some placement matches played by the teams already eliminated. We focus on the first four places, our stylised model contains only a third-place game played between the two losers of the semifinals, similarly to the actual tournaments. Note that there were no playoffs for the 5-8th place in the \href{https://en.wikipedia.org/wiki/2013_World_Men's_Handball_Championship}{2013} and \href{https://en.wikipedia.org/wiki/2017_World_Men's_Handball_Championship}{2017} World Men's Handball Championships.

\begin{center}
\begin{threeparttable}[ht]
\caption{Tournament formats of the \href{https://en.wikipedia.org/wiki/IHF_World_Men\%27s_Handball_Championship}{IHF World \\ Men's Handball Championships} with $24$ teams}
\centering
\label{Table1}
\rowcolors{1}{}{gray!20}
\begin{tabularx}{0.9\linewidth}{Ll ccc ccc} \toprule \hiderowcolors
          &       & \multicolumn{3}{c}{Preliminary round} & \multicolumn{3}{c}{Main round} \\
    Format & Year(s) of application & Gr.   & Teams & Q     & Gr.   & Teams & Q \\ \midrule \showrowcolors
    $KO$    & \href{https://en.wikipedia.org/wiki/1995_World_Men's_Handball_Championship}{1995}-\href{https://en.wikipedia.org/wiki/2001_World_Men's_Handball_Championship}{2001}, \href{https://en.wikipedia.org/wiki/2013_World_Men's_Handball_Championship}{2013}-\href{https://en.wikipedia.org/wiki/2017_World_Men's_Handball_Championship}{2017} & 4     & 6     & 4     & \multicolumn{3}{c}{Knockout} \\
    $G64$   & \href{https://en.wikipedia.org/wiki/2003_World_Men's_Handball_Championship}{2003}  & 4     & 6     & 4     & 4     & 4     & 1 \\
    $G66$   & \href{https://en.wikipedia.org/wiki/2005_World_Men\%27s_Handball_Championship}{2005}, \href{https://en.wikipedia.org/wiki/2009_World_Men\%27s_Handball_Championship}{2009}-\href{https://en.wikipedia.org/wiki/2011_World_Men\%27s_Handball_Championship}{2011}, \href{https://en.wikipedia.org/wiki/2019_World_Men\%27s_Handball_Championship}{2019}-- & 4     & 6     & 3     & 2     & 6     & 2 \\
    $G46$   & \href{https://en.wikipedia.org/wiki/2007_World_Men's_Handball_Championship}{2007}  & 6     & 4     & 2     & 2     & 6     & 4 \\ \hline
\end{tabularx}

\begin{tablenotes}[flushleft]
\item
\footnotesize{Notes: Gr. = Number of groups in the preliminary and main round, respectively; Teams = Number of teams in each group of the preliminary and main round, respectively; Q = Number of teams qualified from each group of the preliminary and main round, respectively}
\end{tablenotes}
\end{threeparttable}
\end{center}

In the following, the designs that have been used recently in the World Men's Handball Championships are presented.
Table~\ref{Table1} and Figures~\ref{Fig_A1}-\ref{Fig_A4} of the Appendix provide an overview of them.

\subsubsection{One group stage with \texorpdfstring{$6$}{6} teams per group (\texorpdfstring{$KO$}{KO})} \label{Sec211}

This design, presented in Figure~\ref{Fig_A1}, has been used in the World Men's Handball Championships between \href{https://en.wikipedia.org/wiki/1995_World_Men's_Handball_Championship}{1995} and \href{https://en.wikipedia.org/wiki/2001_World_Men's_Handball_Championship}{2001} as well as between \href{https://en.wikipedia.org/wiki/2013_World_Men's_Handball_Championship}{2013} and \href{https://en.wikipedia.org/wiki/2017_World_Men's_Handball_Championship}{2017}.
It contains one group stage with four groups of six teams each such that the top four teams qualify for the round of 16 (see~Figure~\ref{Fig_A1a}), where a standard knockout stage starts (see~Figure~\ref{Fig_A1b}). 

\subsubsection{Two group stages with \texorpdfstring{$6$}{6} and \texorpdfstring{$4$}{4} teams per group (\texorpdfstring{$G64$}{G64})} \label{Sec212}

This design, presented in Figure~\ref{Fig_A2}, has been used in the \href{https://en.wikipedia.org/wiki/2001_World_Men's_Handball_Championship}{2003 World Men's Handball Championship}, hosted by Portugal.
It contains two group stages (see~Figure~\ref{Fig_A2a}). The preliminary round consists of four groups of six teams each such that the top four teams qualify for the main round. The main round consists of four groups of four teams each such that two teams in each main round group are from the same preliminary round group, the first and the third, or the second and the fourth. Therefore, all teams play two further matches in the main round. Only the group winners of main round groups qualify for the semifinals in the knockout stage (see~Figure~\ref{Fig_A2b}).

\subsubsection{Two group stages with \texorpdfstring{$6$}{6} and \texorpdfstring{$6$}{6} teams per group (\texorpdfstring{$G66$}{G66})} \label{Sec213}

This design, presented in Figure~\ref{Fig_A3}, has been used first in the \href{https://en.wikipedia.org/wiki/2005_World_Men's_Handball_Championship}{2005 World Men's Handball Championship} and has been applied in \href{https://en.wikipedia.org/wiki/2009_World_Men's_Handball_Championship}{2009}, \href{https://en.wikipedia.org/wiki/2011_World_Men's_Handball_Championship}{2011}, and \href{https://en.wikipedia.org/wiki/2019_World_Men's_Handball_Championship}{2019}.
It contains two group stages (see~Figure~\ref{Fig_A3a}). The preliminary round consists of four groups of six teams each such that the top three teams qualify for the main round. The main round consists of two groups of six teams, each created from two preliminary round groups. Therefore, all teams play three further matches in the main round. The top two teams of every main round group advance to the semifinals in the knockout stage (see~Figure~\ref{Fig_A3b}).

\subsubsection{Two group stages with \texorpdfstring{$4$}{4} and \texorpdfstring{$6$}{6} teams per group (\texorpdfstring{$G46$}{G46})} \label{Sec214}

This design, presented in Figure~\ref{Fig_A4}, has been used in the \href{https://en.wikipedia.org/wiki/2007_World_Men's_Handball_Championship}{2007 World Men's Handball Championship}, hosted by Germany.
It contains two group stages (see~Figure~\ref{Fig_A4a}). Teams are drawn into six groups of four teams each in the preliminary round such that the top two teams proceed to the main round. The main round consists of two groups, each created from three preliminary round groups. Therefore, all teams play four matches in the main round. Four teams of a main round group advance to the quarterfinals in the knockout stage (see~Figure~\ref{Fig_A4b}).

\subsubsection{Round-robin (\texorpdfstring{$RR$}{RR})} \label{Sec215}

While the $24$ competitors have never played a round-robin tournament, we use this basic format as a reference.

\subsubsection{Seeding policy} \label{Sec216}

Seeding plays an important role in knockout tournaments \citep{Hwang1982, Schwenk2000, Marchand2002, GrohMoldovanuSelaSunde2012, Karpov2016, DagaevSuzdaltsev2018, Karpov2018}. It is not an issue in our case since the knockout stage of all formats is immediately determined by the previous group stage (see Figures~\ref{Fig_A1}-\ref{Fig_A4}).
On the other hand, all participants should be drawn into groups before the start of the tournament, and this policy may influence the outcome, too \citep{Guyon2015, DagaevRudyak2019, Guyon2018a, LalienaLopez2018}.

In the recent World Men's Handball Championships, the pots were determined on the basis of geography and other aspects such as qualification results. For example, in the \href{https://en.wikipedia.org/wiki/2009_World_Men's_Handball_Championship#Seeding}{2009} tournament, Pot~1 contained the host (Croatia), the defending World Champions (Germany), the champions of Europe (Denmark), and the third-placed team of the recent European Championship (France), where Croatia and Germany were the second- and fourth-placed teams, respectively.

We consider two variants of each tournament design called \emph{seeded} and \emph{unseeded}.
In the seeded version, the preliminary round groups are seeded such that in the case of $k$ groups ($k=6$ for design $G46$ and $k=4$ otherwise), the strongest $k$ teams are placed in Pot 1, the next strongest $k$ teams in Pot 2, and so on.
Unseeded version applies fully random seeding. In this case, some strong teams, allocated in a harsh group, may have more difficulty in qualifying than weaker teams allocated in an easier group, which is inefficient and can be regarded as unfair.

Naturally, there is no need to seed the teams in the reference format $RR$.

\subsection{Tournament metrics} \label{Sec22}

Following the literature \citep{HorenRiezman1985, ScarfYusofBilbao2009, DagaevRudyak2019}, the following tournament success measures have been chosen:
\begin{itemize}
\item
the probability that one of the best $p$ teams wins the tournament;
\item
the probability that at least one of the best $p$ teams plays in the final;
\item
the average pre-tournament rank of the winner, the second-, the third- and the fourth-placed teams;
\item
the expected quality of the final (the sum of the finalists' pre-tournament ranks);
\item
the expected competitive balance of the final (the difference between the finalists' pre-tournament ranks).
\end{itemize}
We focus only on the first four places because there was a third place game in all World Men's Handball Championships since 1995, however, other placement matches were organised arbitrarily.

\subsection{Simulation procedure} \label{Sec23}

Given the design and a prediction model for match results, we are able to simulate a complete tournament repeatedly and obtain estimates of any metrics of interest.

\subsubsection{Playing abilities} \label{Sec231}

The probability with which a given team would beat another team is fixed \emph{a priori}.
We have chosen a generalised version of \citet{Jackson1993}'s model for this purpose:
\begin{equation} \label{eq1}
p_{ij} = \frac{1}{1+ \left[ (i+\beta)/(j+\beta) \right]^{\alpha}},
\end{equation}
where $p_{ij}$ is the probability that team $i$ defeats team $j$, $\alpha, \beta \geq 0$ are parameters and $1 \leq i,j \leq 24$ is the \emph{identifier} of the teams.
The model was used by \citet{Jackson1993} and \citet{Marchand2002} with $\beta = 0$.
The role of this novel parameter $\beta$ is to lessen the sharp increase of winning probabilities for the strongest teams. The function of $\alpha$ is similar to the original model, its smaller or larger values reflect situations where there is a smaller or larger dispersion in the teams' strengths, respectively.

Stationarity and independence of the probability that team $i$ beats team $j$ is assumed, it does not change throughout the tournament and is independent of the previous results. While in practice they are dynamic and changing probabilities are expected to alter the outcome of the tournament on a single occasion, it seems to be reasonable that stationary probabilities are good approximations of long-run averages \citep{McGarrySchutz1997}.

\begin{figure}[ht]
\centering
\caption[The probability that team $i$ beats its opponent]{The probability that team $i$ beats its opponent (baseline, $\alpha=4$, $\beta=24$)}
\label{Fig1}

\begin{tikzpicture}
\begin{axis}[width=\textwidth, 
height=0.5\textwidth,
xmin=1,
xmax=24,
ymin=0,
ymax=1,
ymajorgrids,
xlabel = Pre-tournament rank of the opponent,
tick label style = {/pgf/number format/fixed},
legend entries = {Team 1 $\qquad$,Team 2 $\qquad$,Team 7 $\qquad$,Team 13},
legend style = {at={(0.5,-0.2)},anchor = north,legend columns = 4}
]
\addplot[blue,smooth,very thick,dashdotdotted] coordinates {
(1,0.5)
(2,0.539140468215587)
(3,0.576358959120063)
(4,0.611427053729256)
(5,0.644209067078602)
(6,0.674648620510151)
(7,0.702753727515816)
(8,0.728582039617816)
(9,0.752227337483334)
(10,0.773807862482129)
(11,0.793456708526107)
(12,0.811314238293996)
(13,0.827522335443614)
(14,0.842220230466511)
(15,0.855541617697201)
(16,0.867612793899598)
(17,0.878551579319149)
(18,0.88846681957479)
(19,0.897458304741974)
(20,0.90561697683898)
(21,0.91302532702477)
(22,0.919757908711872)
(23,0.925881912739033)
(24,0.93145776631542)
};
\addplot[black,smooth,very thick,loosely dotted] coordinates {
(1,0.460859531784413)
(2,0.5)
(3,0.537668817917944)
(4,0.573570031503352)
(5,0.607495595903654)
(6,0.63931755613366)
(7,0.668977187201421)
(8,0.696472788718025)
(9,0.721847443875057)
(10,0.745177637800896)
(11,0.766563257783379)
(12,0.786119202917543)
(13,0.80396862132084)
(14,0.820237660653819)
(15,0.835051546391753)
(16,0.848531774863307)
(17,0.860794209222365)
(18,0.871947884254983)
(19,0.882094351661615)
(20,0.891327425547054)
(21,0.899733214908457)
(22,0.907390354148157)
(23,0.914370363319209)
(24,0.92073808684648)
};
\addplot[ForestGreen,smooth,very thick,loosely dashed] coordinates {
(1,0.297246272484184)
(2,0.331022812798579)
(3,0.365261085856538)
(4,0.399600306076609)
(5,0.433701332227947)
(6,0.467257102740607)
(7,0.5)
(8,0.531706097620959)
(9,0.562196542972028)
(10,0.591336531470797)
(11,0.619032434515083)
(12,0.645227661855676)
(13,0.669897793959428)
(14,0.693045435222427)
(15,0.714695136983381)
(16,0.734888637100221)
(17,0.753680571373399)
(18,0.77113473699184)
(19,0.787320931059465)
(20,0.802312347095235)
(21,0.816183486705999)
(22,0.829008529382739)
(23,0.840860097580612)
(24,0.851808354288562)
};
\addplot[red,smooth,very thick] coordinates {
(1,0.172477664556386)
(2,0.19603137867916)
(3,0.220918090357424)
(4,0.24696713338104)
(5,0.273986787229773)
(6,0.301770273839758)
(7,0.330102206040572)
(8,0.358765089024432)
(9,0.387545497146809)
(10,0.416239604023925)
(11,0.444657824229447)
(12,0.472628417596265)
(13,0.5)
(14,0.52664298737882)
(15,0.552450065693922)
(16,0.57733582519895)
(17,0.601235722635397)
(18,0.624104542107806)
(19,0.645914518184714)
(20,0.666653267540064)
(21,0.686321652357089)
(22,0.704931673304609)
(23,0.722504464866072)
(24,0.739068442983625)
};
\end{axis}
\end{tikzpicture}

\end{figure}
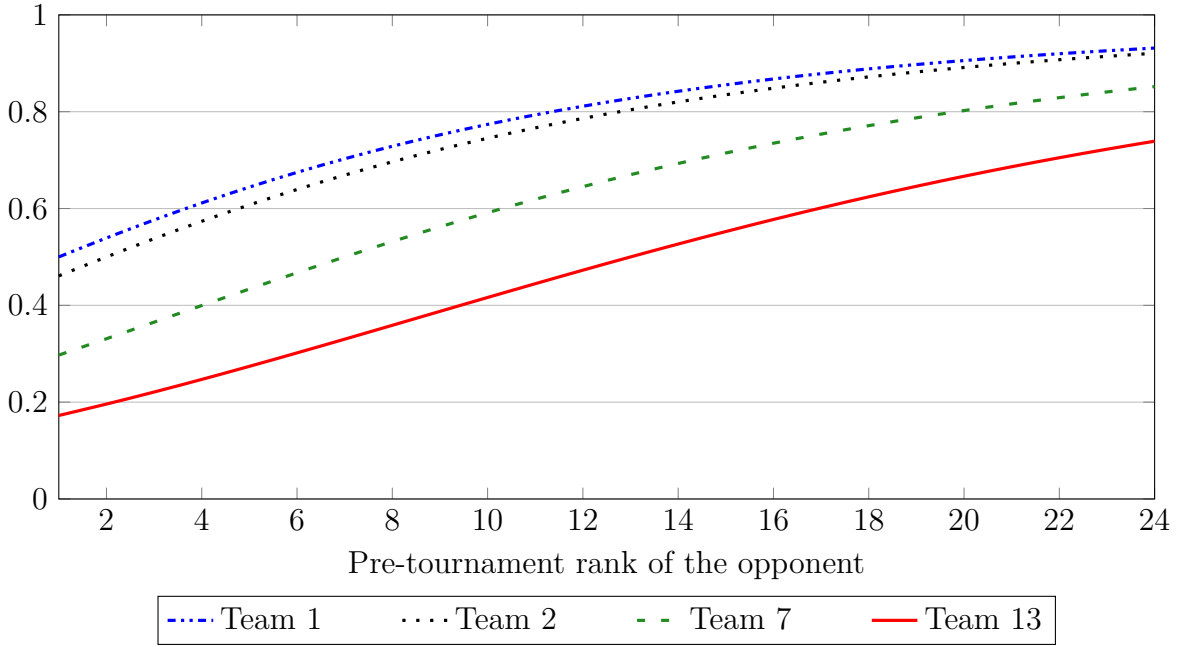


Baseline results are obtained with $\alpha = 4$ and $\beta = 24$ but a robustness check will be provided for both parameters.
Figure~\ref{Fig1} shows the probabilities of beating the opponents for certain teams as derived from formula~\eqref{eq1}. Our choice somewhat follows the idea behind Matrix~I of \citet{McGarrySchutz1997}: adjacent teams are closely matched (team $k-1$ defeats team $k$ with a probability of no more than $0.54$), but the difference between a top team and an underdog is significant (the strongest team has greater than 90\% chance to win against the last five teams).

\subsubsection{Technical details} \label{Sec232}

A handball game may be tied at the end of the regular playing time. If a winner has to be determined, namely, in the knockout stage of a tournament, it is followed by the first and (if it is necessary) the second overtime of 10 minutes, and the match is finally decided with penalty throws.
It is a less frequent event than a draw in football, for example, in the \href{https://en.wikipedia.org/wiki/2017_World_Men's_Handball_Championship}{2017 World Men's Handball Championship}, which was organised according to format $KO$, there were three draws from the $60$ group matches, and one draw from the $16$ matches of the knockout stage.
Thus, following \citet{McGarrySchutz1997}, draws are not allowed in the simulation. This is not to be confused with ties in the ranking of round-robin groups, resolved in our simulations with an equal-odds `coin toss'.

Every simulation has been run one million times ($N=1{,}000{,}000$) such that two matrices with match outcomes have been generated for each possible pair of opponents in every run because some teams may play two matches against each other (however, it is not possible before the semifinals). After that, these outcomes have been plugged into the competition formats analysed to study the outcome of the tournament: we have recorded the identifier of the first four teams and the teams which play the final in each run. Thus any differences in tournament metrics are solely caused by the designs.

The validity of the simulation procedure has been tested in several ways.
First, a matrix representing equality among all teams ($p_{ij} = 0.5$ for all combinations of $i$ and $j$) has led to, as expected, an outcome where all teams are placed first to fourth equally often.
Second, simulations with a fully deterministic matrix ($p_{ij} = 1$ if $i < j$) have been analysed. It still shows the differences between our tournament designs. For example, in the seeded versions of $G66$ and $G46$, the four best teams are guaranteed to occupy the first four places in their natural order. However, in the seeded $KO$ and $G64$, the two strongest teams can meet in the semifinals with a probability of $1/3$. Regarding the unseeded variants, the worst team that may qualify for the semifinals is the sixth in $G46$, the seventh in $KO$, and the fourteenth in $G64$ and $G66$. 
Finally, some values have been changed in the fully deterministic matrix in order to see whether they function in an expected way.


\section{The comparison of tournament designs} \label{Sec3}

In the following, our findings on the four tournament designs that have been used in the recent World Men's Handball Championships, are reviewed.

\subsection{Match distribution} \label{Sec31}

By looking at the tournament formats, it can be realised that two teams may play at most two times against each other, and this number could be two only if one of these matches is a semifinal, the final or the third-place game.

Each design requires different number of matches:
\begin{itemize}
\item
A round-robin tournament with $24$ teams contains $24 \times 23 / 2 = 276$ games.
\item
Format $KO$ contains $4 \times 6 \times 5 / 2 = 60$ games in the group stage, and $8 + 4 + 2 + 2 = 16$ games in the knockout stage, that is, $76$ in total.
\item
Format $G64$ contains $4 \times 6 \times 5 / 2 = 60$ games in the preliminary round, $4 \times 4 \times 2 / 2 = 16$ games in the main round, and $2 + 2 = 4$ games in the knockout stage, that is, $80$ in total.
\item
Format $G66$ contains $4 \times 6 \times 5 / 2 = 60$ games in the preliminary round, $2 \times 6 \times 3 / 2 = 18$ games in the main round, and $2 + 2 = 4$ games in the knockout stage, that is, $82$ in total.
\item
Format $G46$ contains $6 \times 4 \times 3 / 2 = 36$ games in the preliminary round, $2 \times 6 \times 4 / 2 = 24$ games in the main round, and $4 + 2 + 2 = 8$ games in the knockout stage, that is, $68$ in total.
\end{itemize}

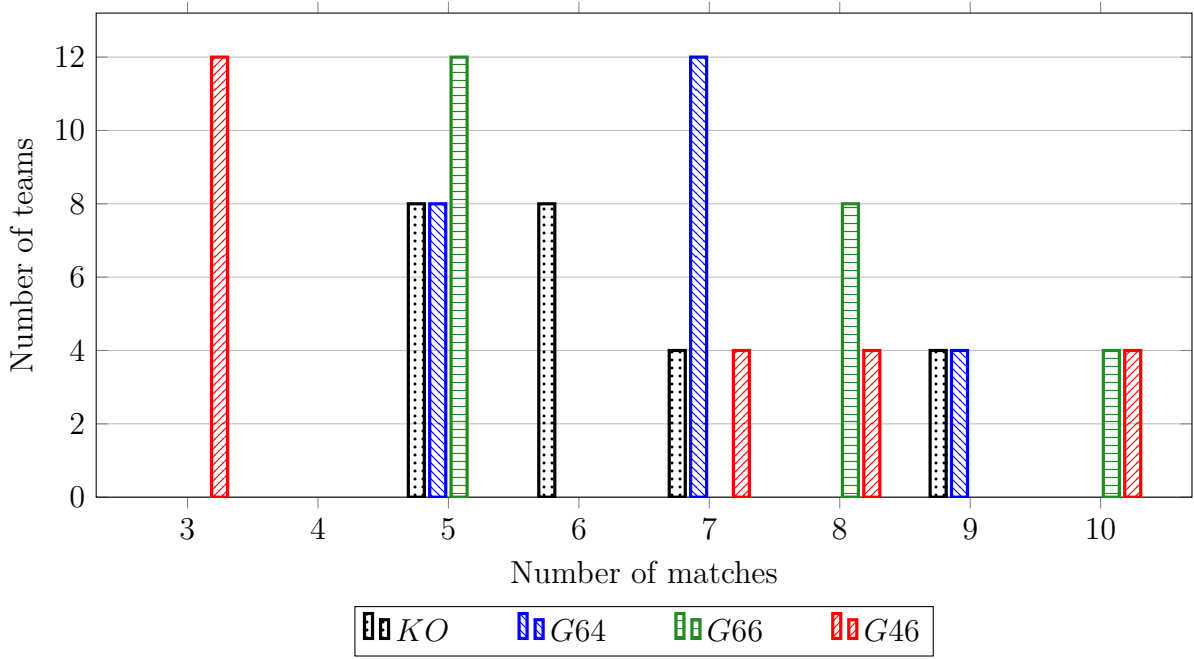
\begin{figure}[ht]
\centering
\caption{The distribution of matches played in different tournament formats}
\label{Fig2}

\begin{tikzpicture}
\begin{axis}[width = \textwidth, 
height = 0.5\textwidth,
xlabel = Number of matches,
ylabel = Number of teams, 
ybar = 2pt,
ymin = 0,
ymajorgrids = true,
bar width = 6pt,
legend entries={$KO \qquad$,$G64 \qquad$,$G66 \qquad$,$G46$},
legend style={at={(0.5,-0.225)},anchor = north,legend columns = 4}
]
\addplot [black, pattern color = black, pattern = dots, very thick] coordinates {
(5,8)
(6,8)
(7,4)
(9,4)
};
\addplot [blue, pattern color = blue, pattern = north west lines, very thick] coordinates {
(5,8)
(7,12)
(9,4)
};
\addplot [ForestGreen, pattern color = ForestGreen, pattern = horizontal lines, very thick] coordinates {
(5,12)
(8,8)
(10,4)
};
\addplot [red, pattern color = red, pattern = north east lines, very thick] coordinates {
(3,12)
(7,4)
(8,4)
(10,4)
};
\end{axis}
\end{tikzpicture}
\end{figure}


Besides the total number of matches, its distribution, presented in Figure~\ref{Fig2}, is also interesting. This reports the number of teams with a given number of matches, for example, under the design $KO$, eight teams play five matches.
In $G46$, half of the teams play only three matches, however, the others play at least seven. For the three remaining designs, the minimum number of games to be played by a team is five. The maximum is nine in $KO$ and $G64$, while ten in $G66$ and $G46$.

\subsection{Main results} \label{Sec32}

\begin{figure}[ht]
\centering
\caption[Dependence of some tournament metrics on the number of iterations]{Dependence of some tournament metrics on the number of iterations \\
\footnotesize{Competition design unseeded $KO$; $\alpha=4$; $\beta=24$}}
\label{Fig3}

\begin{tikzpicture}
\begin{axis}[
name = axis1,
width = 0.495\textwidth, 
height = 0.35\textwidth,
xmin = 1000,
xmax = 10000000,
title = The proportion of tournament wins for \\ the highest pre-tournament ranked team,
title style = {align=center, font=\small},
ymajorgrids,
xlabel = Number of independent runs,
xlabel style = {font=\small},
xmode = log,
tick label style = {/pgf/number format/precision=3},
]
\addplot[red,smooth,very thick] coordinates {
(1000,0.246)
(2500,0.2416)
(5000,0.2478)
(10000,0.2427)
(25000,0.23828)
(50000,0.24666)
(100000,0.24091)
(250000,0.241732)
(500000,0.241316)
(1000000,0.24132)
(2500000,0.2419248)
(5000000,0.2413072)
(10000000,0.2414753)
};
\end{axis}

\begin{axis}[
at = {(axis1.south east)},
xshift = 0.1\textwidth,
width = 0.495\textwidth,
height = 0.35\textwidth,
title style = {align=center},
title = The proportion of finals played by the \\ two highest pre-tournament ranked teams,
title style = {align=center, font=\small},
xmin = 1000,
xmax = 10000000,
ymajorgrids,
xlabel = Number of independent runs,
xlabel style = {font=\small},
xmode = log,
scaled ticks = false,
tick label style = {/pgf/number format/precision=4},
]
\addplot[red,smooth,very thick] coordinates {
(1000,0.089)
(2500,0.0776)
(5000,0.0788)
(10000,0.0772)
(25000,0.0752)
(50000,0.0779)
(100000,0.07762)
(250000,0.07792)
(500000,0.076664)
(1000000,0.07695)
(2500000,0.0773768)
(5000000,0.0772358)
(10000000,0.0772586)
};
\end{axis}
\end{tikzpicture}

\end{figure}
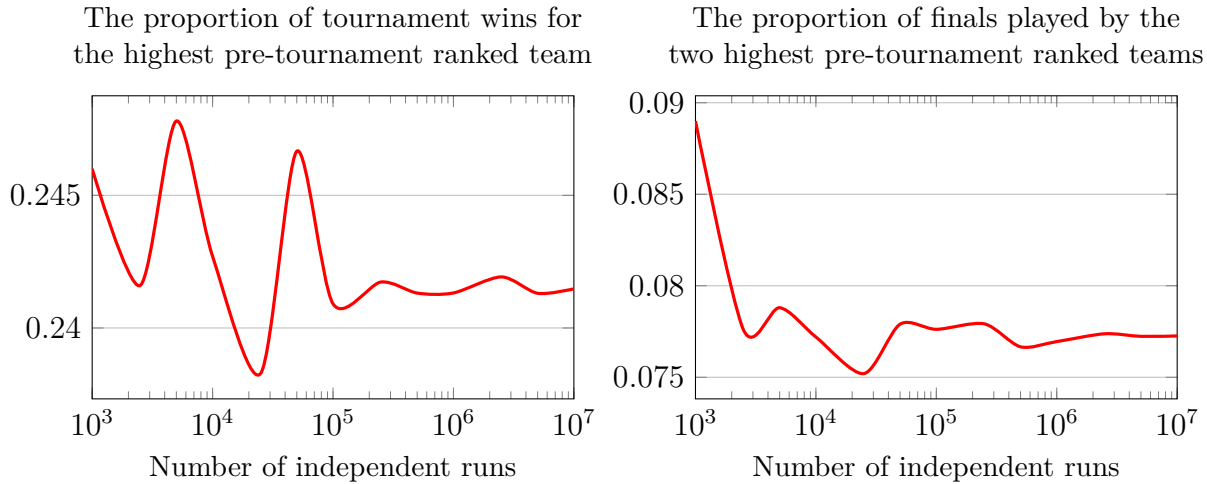

We have tested the simulations with the unseeded variant of tournament design $KO$ for various number of independent runs (Figure~\ref{Fig3}). Since two success measures, the proportion of tournament wins for the highest ranked team, and the proportion of tournament finals between the two highest ranked teams are stable after one million ($10^6$) runs, we have decided to implement all of our following simulations with one million runs.

\begin{figure}[ht]
\centering
\caption[The probability that one of the best $p$ teams \emph{wins} the tournament]{The probability that one of the best $p$ teams \emph{wins} the tournament \\
\footnotesize{Seeded competition designs; $\alpha=4$; $\beta=24$}}
\label{Fig4}

\begin{tikzpicture}
\begin{axis}[width=\textwidth, 
height=0.6\textwidth,
xmin=1,
xmax=12,
ymax=1.05,
ymajorgrids,
xlabel = Value of $p$,
tick label style = {/pgf/number format/fixed},
legend entries={$RR \qquad$,$KO \qquad$,$G64 \qquad$,$G66 \qquad$,$G46$},
legend style={at={(0.5,-0.15)},anchor = north,legend columns = 5}
]
\addplot[brown,smooth,very thick,dashdotted] coordinates {
(1,0.363423)
(2,0.606787)
(3,0.765476)
(4,0.864358)
(5,0.924461)
(6,0.959508)
(7,0.978973)
(8,0.989385)
(9,0.994891)
(10,0.997632)
(11,0.99891)
(12,0.99953)
};
\addplot[black,smooth,very thick,loosely dotted] coordinates {
(1,0.238524)
(2,0.425345)
(3,0.570068)
(4,0.681685)
(5,0.766318)
(6,0.829869)
(7,0.877299)
(8,0.912235)
(9,0.937699)
(10,0.956111)
(11,0.969554)
(12,0.979008)
};
\addplot[blue,smooth,very thick,dashdotdotted] coordinates {
(1,0.239567)
(2,0.426927)
(3,0.571621)
(4,0.683257)
(5,0.766923)
(6,0.829762)
(7,0.877144)
(8,0.912211)
(9,0.936739)
(10,0.955044)
(11,0.968184)
(12,0.977687)
};
\addplot[ForestGreen,smooth,very thick,loosely dashed] coordinates {
(1,0.266219)
(2,0.468575)
(3,0.621891)
(4,0.735944)
(5,0.815352)
(6,0.872477)
(7,0.913262)
(8,0.942161)
(9,0.96095)
(10,0.974018)
(11,0.98294)
(12,0.98903)
};
\addplot[red,smooth,very thick] coordinates {
(1,0.254197)
(2,0.450821)
(3,0.602074)
(4,0.715585)
(5,0.800301)
(6,0.862146)
(7,0.904196)
(8,0.934272)
(9,0.95547)
(10,0.970213)
(11,0.980702)
(12,0.987804)
};
\end{axis}
\end{tikzpicture}

\end{figure}
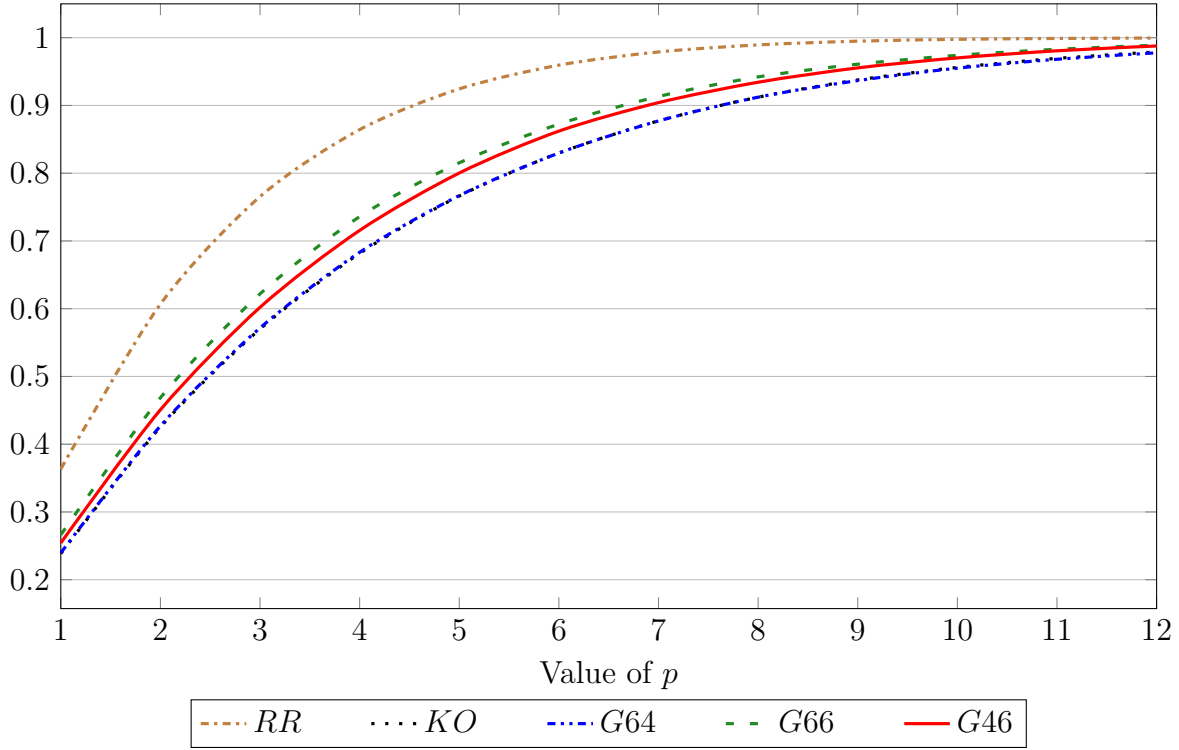


The first tournament metric to be analysed, the probability that one of the best $p$ teams wins the tournament is shown in Figure~\ref{Fig4} for some tournament designs. As expected from the number of matches played in each design (Section~\ref{Sec31}), the round-robin is the format that maximises the probability of winning for the best teams.

\begin{figure}[ht]
\centering
\caption[The probability difference that one of the best $p$ teams \emph{wins} the tournament, compared to a round-robin tournament with $24$ teams]{The probability difference that one of the best $p$ teams \emph{wins} the tournament, \\ compared to a round-robin tournament with $24$ teams ($RR$)}
\label{Fig5}

\begin{tikzpicture}
\begin{axis}[
name = axis1,
width = 0.495\textwidth, 
height = 0.35\textwidth,
xmin = 1,
xmax = 12,
title = {Seeded designs; $\alpha=4$; $\beta=24$},
title style = {align=center, font=\small},
ymajorgrids,
xlabel = Value of $p$,
tick label style = {/pgf/number format/fixed, font=\small},
ymin=-0.21,
ymax=0.01,
]
\addplot[black,smooth,very thick,loosely dotted] coordinates {
(1,-0.124899)
(2,-0.181442)
(3,-0.195408)
(4,-0.182673)
(5,-0.158143)
(6,-0.129639)
(7,-0.101674)
(8,-0.07715)
(9,-0.057192)
(10,-0.041521)
(11,-0.029356)
(12,-0.020522)
};
\addplot[blue,smooth,very thick,dashdotdotted] coordinates {
(1,-0.123856)
(2,-0.17986)
(3,-0.193855)
(4,-0.181101)
(5,-0.157538)
(6,-0.129746)
(7,-0.101829)
(8,-0.077174)
(9,-0.058152)
(10,-0.042588)
(11,-0.030726)
(12,-0.021843)
};
\addplot[ForestGreen,smooth,very thick,loosely dashed] coordinates {
(1,-0.097204)
(2,-0.138212)
(3,-0.143585)
(4,-0.128414)
(5,-0.109109)
(6,-0.087031)
(7,-0.065711)
(8,-0.047224)
(9,-0.033941)
(10,-0.023614)
(11,-0.01597)
(12,-0.0105)
};
\addplot[red,smooth,very thick] coordinates {
(1,-0.109226)
(2,-0.155966)
(3,-0.163402)
(4,-0.148773)
(5,-0.12416)
(6,-0.097362)
(7,-0.074777)
(8,-0.055113)
(9,-0.039421)
(10,-0.027419)
(11,-0.018208)
(12,-0.011726)
};
\end{axis}

\begin{axis}[
at = {(axis1.south east)},
xshift = 0.1\textwidth,
width = 0.495\textwidth,
height = 0.35\textwidth,
xmin = 1,
xmax = 12,
title = {Unseeded designs; $\alpha=4$; $\beta=24$},
title style = {align=center, font=\small},
ymajorgrids,
xlabel = Value of $p$,
tick label style = {/pgf/number format/fixed, font=\small},
legend entries = {$KO \qquad$,$G64 \qquad$,$G66 \qquad$,$G46$},
legend style = {at={(-0.2,-0.3)},anchor=north,legend columns = 4},
ymin=-0.21,
ymax=0.01,
]
\addplot[black,smooth,very thick,loosely dotted] coordinates {
(1,-0.122259)
(2,-0.177359)
(3,-0.190655)
(4,-0.178298)
(5,-0.154972)
(6,-0.127284)
(7,-0.100169)
(8,-0.076367)
(9,-0.057087)
(10,-0.041648)
(11,-0.029874)
(12,-0.021202)
};
\addplot[blue,smooth,very thick,dashdotdotted] coordinates {
(1,-0.124752)
(2,-0.18274)
(3,-0.197258)
(4,-0.185989)
(5,-0.162577)
(6,-0.13454)
(7,-0.107044)
(8,-0.082644)
(9,-0.062428)
(10,-0.046064)
(11,-0.033811)
(12,-0.024492)
};
\addplot[ForestGreen,smooth,very thick,loosely dashed] coordinates {
(1,-0.101319)
(2,-0.145437)
(3,-0.154324)
(4,-0.142387)
(5,-0.121549)
(6,-0.097843)
(7,-0.075341)
(8,-0.05591)
(9,-0.040542)
(10,-0.028713)
(11,-0.019953)
(12,-0.01367)
};
\addplot[red,smooth,very thick] coordinates {
(1,-0.111455)
(2,-0.160442)
(3,-0.170121)
(4,-0.157054)
(5,-0.134104)
(6,-0.108754)
(7,-0.084179)
(8,-0.06314)
(9,-0.046218)
(10,-0.033124)
(11,-0.023369)
(12,-0.016341)
};
\end{axis}
\end{tikzpicture}

\end{figure}


Furthermore, the four designs of the World Men's Handball Championships are almost indistinguishable, therefore it is worth calculating the difference between these formats compared to the reference $RR$, as presented in Figure~\ref{Fig5}. This reveals that design $G66$ is the best from the perspective of its ability to select the strongest teams as the winner, followed by $G46$, while $KO$ and $G64$ perform similarly. Furthermore, seeding has not much effect, with the possible exception of format $KO$: while the seeded variants of $KO$ and $G64$ are almost indistinguishable, $KO$ becomes marginally better with random seeding.

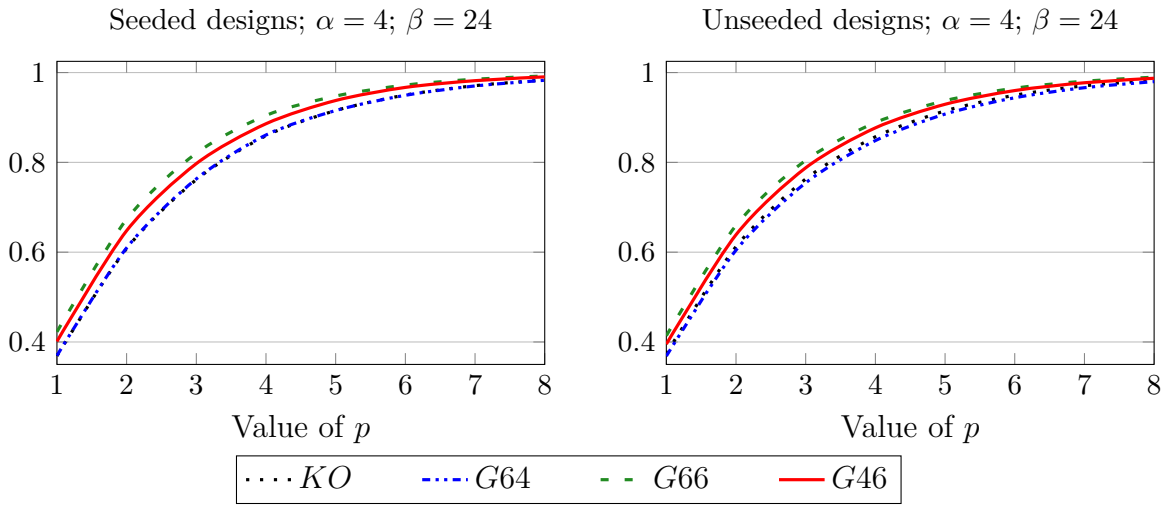
\begin{figure}[ht]
\centering
\caption{The probability that at least one of the best $p$ teams \emph{plays in the final}}
\label{Fig6}

\begin{tikzpicture}
\begin{axis}[
name = axis1,
width = 0.5\textwidth, 
height = 0.35\textwidth,
xmin = 1,
xmax = 8,
xtick = {0,1,2,3,4,5,6,7,8}, 
title = {Seeded designs; $\alpha=4$; $\beta=24$},
title style = {align=center, font=\small},
ymajorgrids,
xlabel = Value of $p$,
tick label style = {/pgf/number format/fixed, font=\small},
ymin = 0.35,
ymax = 1.025,
]
\addplot[black,smooth,very thick,loosely dotted] coordinates {
(1,0.368944)
(2,0.608127)
(3,0.761602)
(4,0.859326)
(5,0.91541)
(6,0.950014)
(7,0.970889)
(8,0.983574)
};
\addplot[blue,smooth,very thick,dashdotdotted] coordinates {
(1,0.369533)
(2,0.609534)
(3,0.76316)
(4,0.860973)
(5,0.915454)
(6,0.949414)
(7,0.970216)
(8,0.982866)
};
\addplot[ForestGreen,smooth,very thick,loosely dashed] coordinates {
(1,0.422527)
(2,0.673332)
(3,0.820289)
(4,0.904576)
(5,0.947755)
(6,0.972133)
(7,0.985269)
(8,0.992351)
};
\addplot[red,smooth,very thick] coordinates {
(1,0.401174)
(2,0.647845)
(3,0.797214)
(4,0.886046)
(5,0.937806)
(6,0.96698)
(7,0.981978)
(8,0.990452)
};
\end{axis}

\begin{axis}[
at = {(axis1.south east)},
xshift = 0.1\textwidth,
width = 0.5\textwidth,
height = 0.35\textwidth,
xmin = 1,
xmax = 8,
xtick = {0,1,2,3,4,5,6,7,8}, 
title = {Unseeded designs; $\alpha=4$; $\beta=24$},
title style = {align=center, font=\small},
ymajorgrids,
xlabel = Value of $p$,
tick label style = {/pgf/number format/fixed, font=\small},
legend entries = {$KO \qquad$,$G64 \qquad$,$G66 \qquad$,$G46$},
legend style = {at={(-0.2,-0.3)},anchor=north,legend columns = 4},
ymin = 0.35,
ymax = 1.025,
]
\addplot[black,smooth,very thick,loosely dotted] coordinates {
(1,0.373832)
(2,0.612286)
(3,0.763008)
(4,0.857009)
(5,0.914575)
(6,0.949526)
(7,0.970603)
(8,0.982761)
};
\addplot[blue,smooth,very thick,dashdotdotted] coordinates {
(1,0.368552)
(2,0.604556)
(3,0.754386)
(4,0.848533)
(5,0.907561)
(6,0.943899)
(7,0.966344)
(8,0.979818)
};
\addplot[ForestGreen,smooth,very thick,loosely dashed] coordinates {
(1,0.413881)
(2,0.659625)
(3,0.80449)
(4,0.88865)
(5,0.937442)
(6,0.965224)
(7,0.98087)
(8,0.989715)
};
\addplot[red,smooth,very thick] coordinates {
(1,0.395171)
(2,0.638827)
(3,0.787893)
(4,0.876857)
(5,0.929362)
(6,0.959944)
(7,0.977335)
(8,0.987329)
};
\end{axis}
\end{tikzpicture}

\end{figure}


The same pattern is attested for our second tournament metric, the probability that at least one of the best $p$ teams plays in the final (Figure~\ref{Fig6}). 

\begin{table}[ht]
\footnotesize{ 
\centering
\caption[Estimates of some tournament metrics for all designs (baseline model)]{Estimates of some tournament metrics for all designs (baseline model) \\
\footnotesize{1 million simulations for each version; $\alpha = 4$; $\beta=24$ \\
S = seeded version of the design; R = unseeded (random) version of the design}}
\label{Table2}
\rowcolors{1}{gray!20}{}
\centerline{
    \begin{tabularx}{1.1\textwidth}{m{3.5cm} CCC CCC CCC} \toprule \hiderowcolors
    & $RR$    & $KO/S$ & $KO/R$ & $G64/S$ & $G64/R$ & $G66/S$ & $G66/R$ & $G46/S$ & $G46/R$ \\ \midrule \showrowcolors
    Min. games & 23 & 5     & 5     & 5     & 5     & 5     & 5     & 3     & 3 \\
    Max. games & 23 & 9     & 9     & 9     & 9     & 10    & 10    & 10    & 10 \\
    Total games & 276 & 76    & 76    & 80    & 80    & 82    & 82    & 68    & 68 \\ \hline
    Average rank of \#1 & 2.56  & 3.90  & 3.88  & 3.90  & 3.96  & 3.48  & 3.58  & 3.60  & 3.70 \\
    Average rank of \#2 & 3.31  & 5.70  & 5.69  & 5.75  & 5.86  & 4.88  & 5.11  & 5.07  & 5.31 \\
    Average rank of \#3 & 4.03  & 5.81  & 5.86  & 5.74  & 6.11  & 4.94  & 5.37  & 5.19  & 5.45 \\
    Average rank of \#4 & 4.79  & 8.63  & 8.74  & 8.63  & 9.19  & 7.16  & 7.87  & 7.47  & 8.01 \\
    Proportion of wins for the highest ranked & 0.36  & 0.24  & 0.24  & 0.24  & 0.24  & 0.27  & 0.26  & 0.25  & 0.25 \\
    Expected quality \newline of the final & ---   & 9.60  & 9.57  & 9.65  & 9.82  & 8.35  & 8.69  & 8.68  & 9.01 \\
    Expected competitive \newline balance of the final & ---   & 4.40  & 4.37  & 4.45  & 4.51  & 3.78  & 3.94  & 3.87  & 4.09 \\ \hline
    \end{tabularx}
}
}
\end{table}

The remaining outcome characteristics for the nine tournament designs are summarised in Table~\ref{Table2}.
As it has already been mentioned in the Introduction, a format is said to be more efficacious if the average pre-tournament rank of the team finishing in the $p$th place ($p = 1,2,3,4$) is smaller, as well as the expected quality and the expected competitive balance of the final is lower (more favourable).

While the round-robin design shows the best performance in selecting the highest pre-tournament ranked teams as the winner, it requires a large number of matches, and only the other formats can be applied in practice.
Among the seeded variants, $G66$ is the most efficacious, followed by $G46$, while the order of $KO$ and $G64$ remains undecided, although the former has a marginal advantage. This order holds for all criteria of efficacy considered here, although they can be conflicting.

Eliminating the seeding procedure changes the metrics according to our expectations, for example, the unseeded $G66$ is approximately at the same level as the seeded $G46$, whereas the unseeded $G46$ is still more efficacious than the $KO$. An interesting observation -- perhaps a kind of puzzle -- is that the performance of the design $KO$, where the knockout stage plays the greatest role, is essentially not influenced by seeding.

The metrics of seeded $G64$ may refer to a flaw of this format because the average pre-tournament ranking of the bronze medallist is not substantially greater than the average ranking of the silver medallist. The unexpected phenomenon is perhaps caused by its strange knockout stage, where only the group winners of the main round compete.

Focusing on the averages of our success measures may mask some variance behind them. For instance, the same expected pre-tournament rank of the winner can be obtained if a format gives a higher probability for the top and the bottom teams, but harm the middle teams.
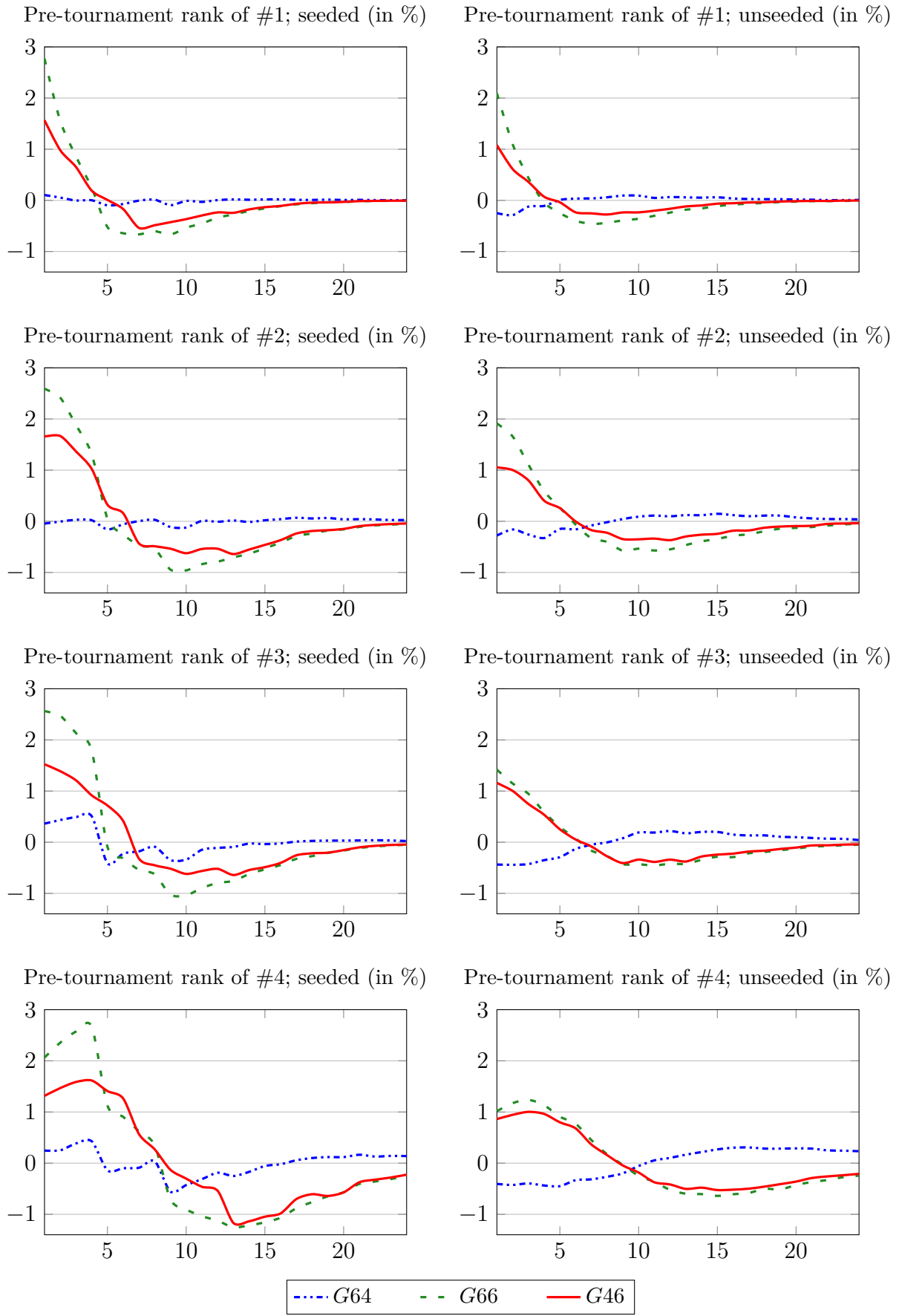
Therefore, Figure~\ref{Fig7} plots the probability difference of reaching the first four places as a function of the pre-tournament rank, compared to competition design $KO$. For example, the probability of the strongest team winning the championship is more than $2.5$\% higher under format $G66$ than under format $KO$, a more than 10\% increase.

The graph reinforces that the difference between seeded $KO$ and $G64$ is negligible, especially in the allocation of the first two places. On the other hand, designs $G66$ and $G46$ are preferred only by the four or five strongest teams. In addition, the lines do not converge to zero even for the weakest teams in the case of the fourth place (\#4) because the presence of two well-constructed subsequent group stages is effective against the occasional emergence of underdogs in the semifinals.

It is reasonable to assume that if a given design is more efficacious than another in both its seeded and unseeded variants, then it remains more efficacious in real-life when the actual allocation of the teams into pots is somewhere between these two extreme cases.

Naturally, all results should be considered with respect to the number of games played. It is the smallest, $68$ for the design $G46$, so its second-place according to efficacy has a favourable message for the organisers: there exists no clear trade-off between efficacy and the number of matches. This is in contrast to the intuition and the conclusions of many tournament design papers. For example, the performance of soccer league formats highly depends on the total number of matches played \citep{LasekGagolewski2018}.

The underlying reason is probably that half of the participating teams play only three games in $G46$, which seems to be enough to determine the competitors with the greatest chance to win the tournament. In addition, this is the only design containing quarterfinals after two group stages.
The remaining three formats are closer to each other from this point of view, all teams play at least five matches, and the total number of games is between $76$ and $82$. This fact also shows that $G64$ is a misaligned design because of the relatively high number of matches could not reduce outcome uncertainty.

\subsection{Sensitivity analysis} \label{Sec42}

Following \citet{ScarfYusofBilbao2009}, the robustness of the results is addressed by calculating our metrics for more and less competitive tournaments than the baseline version. It is achieved in two ways, by changing: the parameter $\alpha$ in formula \eqref{eq1} from its original value of $4$ to $3$ (more competitive) and $5$ (less competitive); and the parameter $\beta$ in formula \eqref{eq1} from its original value of $24$ to $18$ (less competitive) and $36$ (more competitive).

Figure~\ref{Fig8} reproduces Figure~\ref{Fig5} for these cases. It can be seen that the ranking of the competition designs by their ability to select the best teams as the winner remains unchanged as $G66$ is better than $G46$, which outperforms $KO$ and $G64$. The seeded $KO$ and $G64$ formats are almost indistinguishable, but the former outperforms the latter from this point of view without seeding. The advantage of an ideal round-robin tournament becomes more significant if competitive balance is smaller, that is, the outcome of the matches is more difficult to forecast.

Figure~\ref{Fig9} reinforces that seeding has not much influence on tournament outcomes, and, while the actual differences among the four designs are modest (at least compared to the round-robin format), they are robust with respect to the distribution of teams' strength. 

\begin{table}[ht!]
\footnotesize{ 
\centering
\caption[Sensitivity analysis for parameter $\alpha$ -- Estimates of some tournament metrics]{Sensitivity analysis for parameter $\alpha$ -- Estimates of some tournament metrics \\
\footnotesize{1 million simulations for each version \\
S = seeded version of the design; R = unseeded (random) version of the design}}
\label{Table3}

\begin{subtable}{\textwidth}
\caption{$\boldsymbol{\alpha=3}$; $\beta=24$ (more competitive)}
\rowcolors{1}{gray!20}{}
\centerline{
    \begin{tabularx}{1.1\textwidth}{m{3.5cm} CCC CCC CCC} \toprule \hiderowcolors
    & $RR$    & $KO/S$ & $KO/R$ & $G64/S$ & $G64/R$ & $G66/S$ & $G66/R$ & $G46/S$ & $G46/R$ \\ \midrule \showrowcolors    
    Average rank of \#1 & 3.00  & 4.78  & 4.78  & 4.81  & 4.89  & 4.26  & 4.40  & 4.41  & 4.55 \\
    Average rank of \#2 & 3.82  & 6.73  & 6.77  & 6.82  & 6.96  & 5.87  & 6.11  & 6.05  & 6.36 \\
    Average rank of \#3 & 4.58  & 6.84  & 6.89  & 6.78  & 7.16  & 5.90  & 6.33  & 6.15  & 6.48 \\
    Average rank of \#4 & 5.35  & 9.58  & 9.72  & 9.61  & 10.08 & 8.24  & 8.86  & 8.53  & 9.08 \\
    Proportion of wins for the highest ranked & 0.31  & 0.20  & 0.20  & 0.20  & 0.19  & 0.22  & 0.22  & 0.21  & 0.21 \\
    Expected quality \newline of the final & ---   & 11.51 & 11.55 & 11.63 & 11.86 & 10.13 & 10.51 & 10.46 & 10.91 \\
    Expected competitive \newline balance of the final & ---   & 5.29  & 5.30  & 5.38  & 5.45  & 4.65  & 4.84  & 4.75  & 5.00 \\ \bottomrule
    \end{tabularx}
}
\end{subtable}

\vspace{0.5cm}
\begin{subtable}{\textwidth}
\caption{$\boldsymbol{\alpha=5}$; $\beta=24$ (less competitive)}
\rowcolors{1}{gray!20}{}
\centerline{
    \begin{tabularx}{1.1\textwidth}{m{3.5cm} CCC CCC CCC} \toprule \hiderowcolors
    & $RR$    & $KO/S$ & $KO/R$ & $G64/S$ & $G64/R$ & $G66/S$ & $G66/R$ & $G46/S$ & $G46/R$ \\ \midrule \showrowcolors
    Average rank of \#1 & 2.31  & 3.34  & 3.33  & 3.34  & 3.38  & 3.00  & 3.08  & 3.13  & 3.18 \\
    Average rank of \#2 & 3.02  & 5.02  & 4.98  & 5.04  & 5.12  & 4.26  & 4.46  & 4.48  & 4.63 \\
    Average rank of \#3 & 3.73  & 5.11  & 5.17  & 5.03  & 5.40  & 4.34  & 4.77  & 4.60  & 4.78 \\
    Average rank of \#4 & 4.50  & 7.85  & 7.96  & 7.82  & 8.44  & 6.38  & 7.16  & 6.73  & 7.21 \\
    Proportion of wins for the highest ranked & 0.40  & 0.28  & 0.28  & 0.28  & 0.28  & 0.31  & 0.30  & 0.29  & 0.29 \\
    Expected quality \newline of the final & ---   & 8.36  & 8.30  & 8.38  & 8.50  & 7.27  & 7.55  & 7.61  & 7.81 \\
    Expected competitive \newline balance of the final & ---   & 3.80  & 3.75  & 3.82  & 3.86  & 3.22  & 3.37  & 3.35  & 3.49 \\ \bottomrule
    \end{tabularx}
}
\end{subtable}
}
\end{table}

Further tournament characteristics are summarised in Tables~\ref{Table3} and \ref{Table4}.
Our conclusions do not change significantly, although the unseeded $KO$ is clearly worse than its seeded variant if $\alpha=3$ or $\beta=36$ when the teams' abilities are more similar. The final of the seeded $KO$ is more exciting with higher quality and outcome uncertainty than the final of the seeded $G64$ in these cases, too.
A possible flaw of design $G64$ under seeding (the average pre-tournament rank of the third-placed team is close to the average rank of the second-placed) can be observed as before, too.

\begin{table}[ht!]
\footnotesize{ 
\centering
\caption[Sensitivity analysis for parameter $\beta$ -- Estimates of some tournament metrics]{Sensitivity analysis for parameter $\beta$ -- Estimates of some tournament metrics \\
\footnotesize{1 million simulations for each version \\
S = seeded version of the design; R = unseeded (random) version of the design}}
\label{Table4}

\begin{subtable}{\textwidth}
\caption{$\alpha=4$; $\boldsymbol{\beta=18}$ (less competitive)}
\rowcolors{1}{gray!20}{}
\centerline{
    \begin{tabularx}{1.1\textwidth}{m{3.5cm} CCC CCC CCC} \toprule \hiderowcolors
    & $RR$    & $KO/S$ & $KO/R$ & $G64/S$ & $G64/R$ & $G66/S$ & $G66/R$ & $G46/S$ & $G46/R$ \\ \midrule \showrowcolors
    Average rank of \#1 & 2.27  & 3.33  & 3.31  & 3.32  & 3.36  & 2.98  & 3.06  & 3.11  & 3.16 \\
    Average rank of \#2 & 2.99  & 5.05  & 5.01  & 5.07  & 5.17  & 4.27  & 4.48  & 4.48  & 4.66 \\
    Average rank of \#3 & 3.72  & 5.15  & 5.21  & 5.07  & 5.44  & 4.36  & 4.79  & 4.61  & 4.82 \\
    Average rank of \#4 & 4.51  & 7.95  & 8.07  & 7.93  & 8.57  & 6.45  & 7.26  & 6.82  & 7.30 \\
    Proportion of wins for the highest ranked & 0.41  & 0.28  & 0.29  & 0.28  & 0.28  & 0.31  & 0.31  & 0.30  & 0.30 \\
    Expected quality \newline of the final & ---   & 8.37  & 8.32  & 8.40  & 8.53  & 7.25  & 7.54  & 7.59  & 7.82 \\
    Expected competitive \newline balance of the final & ---   & 3.84  & 3.80  & 3.87  & 3.92  & 3.24  & 3.40  & 3.37  & 3.52 \\ \bottomrule
    \end{tabularx}
}
\end{subtable}

\vspace{0.5cm}
\begin{subtable}{\textwidth}
\caption{$\alpha=4$; $\boldsymbol{\beta=36}$ (more competitive)}
\rowcolors{1}{gray!20}{}
\centerline{
    \begin{tabularx}{1.1\textwidth}{m{3.5cm} CCC CCC CCC} \toprule \hiderowcolors
    & $RR$    & $KO/S$ & $KO/R$ & $G64/S$ & $G64/R$ & $G66/S$ & $G66/R$ & $G46/S$ & $G46/R$ \\ \midrule \showrowcolors
    Average rank of \#1 & 3.13  & 4.90  & 4.91  & 4.94  & 5.02  & 4.41  & 4.54  & 4.54  & 4.69 \\
    Average rank of \#2 & 3.94  & 6.82  & 6.85  & 6.90  & 7.04  & 5.96  & 6.21  & 6.14  & 6.45 \\
    Average rank of \#3 & 4.68  & 6.92  & 6.97  & 6.86  & 7.21  & 5.99  & 6.40  & 6.23  & 6.56 \\
    Average rank of \#4 & 5.42  & 9.59  & 9.72  & 9.62  & 10.07 & 8.27  & 8.88  & 8.55  & 9.09 \\
    Proportion of wins for the highest ranked & 0.30  & 0.19  & 0.19  & 0.19  & 0.19  & 0.21  & 0.21  & 0.20  & 0.20 \\
    Expected quality \newline of the final & ---   & 11.72 & 11.76 & 11.84 & 12.06 & 10.37 & 10.75 & 10.68 & 11.14 \\
    Expected competitive \newline balance of the final & ---   & 5.34  & 5.34  & 5.42  & 5.49  & 4.72  & 4.89  & 4.80  & 5.06 \\ \bottomrule
    \end{tabularx}
}
\end{subtable}
}
\end{table}

\section{Discussion} \label{Sec4}

We have compared four tournament formats of recent World Men's Handball Championships. They have been evaluated by Monte-Carlo simulations under two seeding policies, namely, allocating teams perfectly into pots on the basis of their known strength, and a fully random draw of groups.
Our main findings are as follows:
\begin{itemize}
\item
$KO$ (applied from 1995 to 2001 and between 2013 and 2017, see Section~\ref{Sec211}): it is almost insensitive to the seeding rule. While this seems to be a somewhat surprising fact because the knockout phase plays the greatest role in this format, \citet{Marchand2002} provides evidence that the outcome of the standard and random knockout tournaments may not vary as much as one might expect.
\item
$G64$ (applied in 2003, see Section~\ref{Sec212}): it turns out to be a questionable design because of its weak ability to select the best teams despite the relatively high number of matches, and the average pre-tournament rank of the third-placed team is not substantially higher than the average pre-tournament rank of the second-placed team in the seeded variant.
\item
$G66$ (the actual design in 2019, applied in 2005, 2009 and 2011, see Section~\ref{Sec213}): it maximises the association between teams' strength and final position, partially due to the highest number of matches played among the designs considered.
\item
$G46$ (applied in 2007, see Section~\ref{Sec214}): it means a good compromise between efficacy and compactness as the only format with better performance ($G66$) requires a 20\% increase in the number of matches.
\end{itemize}

Our analysis clearly shows that no single best tournament design exists. For example, format $G46$ allows only three matches for certain teams before they are eliminated, which may be regarded as the price for an appropriate selection of top teams. On the other hand, format $KO$ is insensitive to the drawing of groups, therefore this competition structure minimises randomness in a sense by being independent of the seeding policy. In short, we can agree with \citet{ScarfYusofBilbao2009} that one cannot come up with a unique definition of fairness that all would accept.

Nonetheless, the current paper has an important message for the governing bodies of major sports: the obvious conclusion from the intuition and the principle of statistics that a bigger sample lead to better estimates does not necessarily hold in the case of such complex hybrid tournament designs as the comparison of formats $G64$ and $G46$ reveals.

Naturally, all results are based on a particular probabilistic model, which implies certain limitations. However, we have made great effort to minimise this sensitivity by studying a variety of robustness check, and it seems that a wide range of model assumptions are appropriate for comparative purposes \citep{Appleton1995}.

These competition designs have been used in other team tournaments with $24$ participants, too.
The \href{https://en.wikipedia.org/wiki/IHF_World_Women\%27s_Handball_Championship}{IHF World Women's Handball Championship} is organised in every two years since 1993, and has $24$ teams since 1997. Its format has followed the World Men's Handball Championship taken place in the same year, except for \href{https://en.wikipedia.org/wiki/2003_World_Women\%27s_Handball_Championship}{2003} -- when women handball teams competed under design $G66$, while men played in format $G64$, thus no women tournament was organised according to this dubious design --, and for \href{https://en.wikipedia.org/wiki/2011_World_Women\%27s_Handball_Championship}{2011} -- when women national teams competed under design $KO$, while men played in format $G66$.
Similarly to the Men's Championship, the next Women's Championship to be held in \href{https://en.wikipedia.org/wiki/2019_World_Women\%27s_Handball_Championship}{2019}, hosted by Japan, will also use the format $G66$ instead of $KO$ \citep{IHF2018}.

In basketball, the \href{https://en.wikipedia.org/wiki/2006_FIBA_World_Championship}{2006} and the \href{https://en.wikipedia.org/wiki/2010_FIBA_World_Championship}{2010 FIBA World Championship}s as well as the \href{https://en.wikipedia.org/wiki/2014_FIBA_Basketball_World_Cup}{2014 FIBA Basketball World Cup} (the tournament previously known as the FIBA World Championship), the \href{https://en.wikipedia.org/wiki/EuroBasket_2015}{EuroBasket 2015}, and the \href{https://en.wikipedia.org/wiki/EuroBasket_2017}{EuroBasket 2017} applied the design $KO$.
Format $G66$ was used in the \href{https://en.wikipedia.org/wiki/1986_FIBA_World_Championship}{1986 FIBA World Championship}, while the \href{https://en.wikipedia.org/wiki/EuroBasket_2011}{EuroBasket 2011} and the \href{https://en.wikipedia.org/wiki/EuroBasket_2013}{EuroBasket 2013} applied $G66$ with a slight modification that four teams advanced from each of the two main round groups to the quarterfinals (instead of only two to the semifinals).
Finally, the \href{https://en.wikipedia.org/wiki/1978_FIVB_Volleyball_Men's_World_Championship}{1978} and \href{https://en.wikipedia.org/wiki/1978_FIVB_Volleyball_Men's_World_Championship}{1982 FIVB Volleyball Men's World Championship}s were organised in a structure similar to $G46$, but only the two top teams from the two main round groups qualified for the semifinals, while in handball, the second group stage was followed by the quarterfinals (see Figure~\ref{Fig_A4}).
In the view of our computations, perhaps it is not a coincidence that no further use of the strange design $G64$ has been found in practice.

Organisers of team championships are encouraged to consider our results when deciding on the design of future tournaments. For example, the recent change of the World Men's Handball Championship format (from $KO$ to $G66$ between 2017 and 2019) has increased the probability of winning for the best teams as revealed by Figure~\ref{Fig7}.
The choice of tournament design is an especially important issue because it offers perhaps the only way to influence the expected value of certain success measures for sports administrators.

There is a great scope for future research.
First, one can implement a more extensive sensitivity analysis.
Second, as discussed in the Introduction, our simulation is not based on data from real tournaments since it is far from trivial to model handball matches.
Third, other tournament designs or simple modifications of the formats analysed here (recall that a slightly modified variant of structure $G46$ was used in volleyball) can be investigated with the presented methodology.
Finally, further properties of the competition formats are worth examining. For example, it is almost obvious to check that design $KO$ satisfies strategy-proofness, while formats $G64$, $G66$, and $G46$ are incentive incompatible \citep{Csato2019i}. 

\section*{Acknowledgements}
\addcontentsline{toc}{section}{Acknowledgements}
\noindent
This paper could not be written without my father, who has coded the simulations in Python mainly during a Christmas break. \\
We are grateful to \emph{Tam\'as Halm} for reading the manuscript. \\
Four anonymous reviewers provided valuable comments and suggestions on earlier drafts. \\
We are indebted to the \href{https://en.wikipedia.org/wiki/Wikipedia_community}{Wikipedia community} for contributing to our research by collecting and structuring some information used in the paper. \\
The research was supported by OTKA grant K 111797 and by the MTA Premium Postdoctoral Research Program. 

\bibliographystyle{apalike}
\bibliography{All_references}

\clearpage

\section*{Appendix}
\addcontentsline{toc}{section}{Appendix}

\renewcommand\thefigure{A.\arabic{figure}}
\setcounter{figure}{0}

\makeatletter
\renewcommand\p@subfigure{A.\arabic{figure}}
\makeatother

\input{Figures_TD_Appendix}

\input{Figure7_pre_tournament_rank_distribution}

\input{Figure8_winning_probability_difference_SA}

\begin{figure}[ht]
\centering
\caption{Sensitivity analysis -- The probability that \\ at least one of the best $p$ teams \emph{plays in the final}}
\label{Fig9}

\begin{tikzpicture}
\begin{axis}[
name = axis1,
width = 0.5\textwidth, 
height = 0.35\textwidth,
xmin = 1,
xmax = 8,
xtick = {0,1,2,3,4,5,6,7,8}, 
title = {Seeded designs; $\boldsymbol{\alpha=3}$; $\beta=24$},
title style = {align=center, font=\small},
ymajorgrids,
tick label style = {/pgf/number format/fixed},
ymin = 0.29,
ymax = 1.025,
]
\addplot[black,smooth,very thick,loosely dotted] coordinates {
(1,0.309364)
(2,0.527808)
(3,0.680491)
(4,0.787076)
(5,0.856596)
(6,0.904528)
(7,0.936915)
(8,0.959127)
};
\addplot[blue,smooth,very thick,dashdotdotted] coordinates {
(1,0.307814)
(2,0.525941)
(3,0.679408)
(4,0.786795)
(5,0.855329)
(6,0.902881)
(7,0.935532)
(8,0.957682)
};
\addplot[ForestGreen,smooth,very thick,loosely dashed] coordinates {
(1,0.35413)
(2,0.588196)
(3,0.741343)
(4,0.840327)
(5,0.899191)
(6,0.937212)
(7,0.961085)
(8,0.976358)
};
\addplot[red,smooth,very thick] coordinates {
(1,0.338367)
(2,0.567231)
(3,0.719857)
(4,0.82046)
(5,0.886276)
(6,0.929147)
(7,0.95513)
(8,0.972087)
};
\end{axis}

\begin{axis}[
at = {(axis1.south east)},
xshift = 0.1\textwidth,
width = 0.5\textwidth,
height = 0.35\textwidth,
xmin = 1,
xmax = 8,
xtick = {0,1,2,3,4,5,6,7,8}, 
title = {Unseeded designs; $\boldsymbol{\alpha=3}$; $\beta=24$},
title style = {align=center, font=\small},
ymajorgrids,
tick label style = {/pgf/number format/fixed},
ymin = 0.29,
ymax = 1.025,
]
\addplot[black,smooth,very thick,loosely dotted] coordinates {
(1,0.311845)
(2,0.529454)
(3,0.679994)
(4,0.783857)
(5,0.854649)
(6,0.902867)
(7,0.93546)
(8,0.957282)
};
\addplot[blue,smooth,very thick,dashdotdotted] coordinates {
(1,0.306037)
(2,0.520308)
(3,0.669757)
(4,0.773444)
(5,0.84511)
(6,0.894302)
(7,0.928512)
(8,0.95211)
};
\addplot[ForestGreen,smooth,very thick,loosely dashed] coordinates {
(1,0.346284)
(2,0.574937)
(3,0.72547)
(4,0.823013)
(5,0.886638)
(6,0.927654)
(7,0.95392)
(8,0.970872)
};
\addplot[red,smooth,very thick] coordinates {
(1,0.331413)
(2,0.55516)
(3,0.70595)
(4,0.806944)
(5,0.873793)
(6,0.917732)
(7,0.946922)
(8,0.96572)
};
\end{axis}
\end{tikzpicture}

\begin{tikzpicture}
\begin{axis}[
name = axis1,
width = 0.5\textwidth, 
height = 0.35\textwidth,
xmin = 1,
xmax = 8,
xtick = {0,1,2,3,4,5,6,7,8}, 
title = {Seeded designs; $\boldsymbol{\alpha=5}$; $\beta=24$},
title style = {align=center, font=\small},
ymajorgrids,
tick label style = {/pgf/number format/fixed},
ymin = 0.39,
ymax = 1.025,
]
\addplot[black,smooth,very thick,loosely dotted] coordinates {
(1,0.420632)
(2,0.673324)
(3,0.821199)
(4,0.906394)
(5,0.94994)
(6,0.973986)
(7,0.986723)
(8,0.993484)
};
\addplot[blue,smooth,very thick,dashdotdotted] coordinates {
(1,0.420962)
(2,0.673265)
(3,0.821878)
(4,0.907733)
(5,0.949969)
(6,0.973443)
(7,0.986195)
(8,0.993156)
};
\addplot[ForestGreen,smooth,very thick,loosely dashed] coordinates {
(1,0.47757)
(2,0.736411)
(3,0.872957)
(4,0.941683)
(5,0.972048)
(6,0.986998)
(7,0.99422)
(8,0.99758)
};
\addplot[red,smooth,very thick] coordinates {
(1,0.453496)
(2,0.708995)
(3,0.84999)
(4,0.925519)
(5,0.964541)
(6,0.98375)
(7,0.99246)
(8,0.996606)
};
\end{axis}

\begin{axis}[
at = {(axis1.south east)},
xshift = 0.1\textwidth,
width = 0.5\textwidth,
height = 0.35\textwidth,
xmin = 1,
xmax = 8,
xtick = {0,1,2,3,4,5,6,7,8}, 
title = {Unseeded designs; $\boldsymbol{\alpha=5}$; $\beta=24$},
title style = {align=center, font=\small},
ymajorgrids,
tick label style = {/pgf/number format/fixed},
ymin = 0.39,
ymax = 1.025,
]
\addplot[black,smooth,very thick,loosely dotted] coordinates {
(1,0.426444)
(2,0.677424)
(3,0.822025)
(4,0.903777)
(5,0.948773)
(6,0.973189)
(7,0.986292)
(8,0.993127)
};
\addplot[blue,smooth,very thick,dashdotdotted] coordinates {
(1,0.421815)
(2,0.670628)
(3,0.815132)
(4,0.897239)
(5,0.943702)
(6,0.969388)
(7,0.983695)
(8,0.991325)
};
\addplot[ForestGreen,smooth,very thick,loosely dashed] coordinates {
(1,0.470507)
(2,0.723721)
(3,0.858363)
(4,0.928781)
(5,0.964909)
(6,0.982924)
(7,0.99184)
(8,0.996209)
};
\addplot[red,smooth,very thick] coordinates {
(1,0.450207)
(2,0.705006)
(3,0.844444)
(4,0.919709)
(5,0.959323)
(6,0.979785)
(7,0.990068)
(8,0.995224)
};
\end{axis}
\end{tikzpicture}

\begin{tikzpicture}
\begin{axis}[
name = axis1,
width = 0.5\textwidth, 
height = 0.35\textwidth,
xmin = 1,
xmax = 8,
xtick = {0,1,2,3,4,5,6,7,8}, 
title = {Seeded designs; $\alpha=4$; $\boldsymbol{\beta=18}$},
title style = {align=center, font=\small},
ymajorgrids,
tick label style = {/pgf/number format/fixed},
ymin = 0.39,
ymax = 1.025,
]
\addplot[black,smooth,very thick,loosely dotted] coordinates {
(1,0.425298)
(2,0.677641)
(3,0.823913)
(4,0.907563)
(5,0.950369)
(6,0.973804)
(7,0.986408)
(8,0.99324)
};
\addplot[blue,smooth,very thick,dashdotdotted] coordinates {
(1,0.426432)
(2,0.678387)
(3,0.825602)
(4,0.909576)
(5,0.950499)
(6,0.97349)
(7,0.986032)
(8,0.99292)
};
\addplot[ForestGreen,smooth,very thick,loosely dashed] coordinates {
(1,0.48386)
(2,0.742341)
(3,0.876583)
(4,0.943317)
(5,0.972701)
(6,0.98723)
(7,0.994245)
(8,0.997411)
};
\addplot[red,smooth,very thick] coordinates {
(1,0.459099)
(2,0.714574)
(3,0.853924)
(4,0.927279)
(5,0.964988)
(6,0.984105)
(7,0.992529)
(8,0.996455)
};
\end{axis}

\begin{axis}[
at = {(axis1.south east)},
xshift = 0.1\textwidth,
width = 0.5\textwidth,
height = 0.35\textwidth,
xmin = 1,
xmax = 8,
xtick = {0,1,2,3,4,5,6,7,8}, 
title = {Unseeded designs; $\alpha=4$; $\boldsymbol{\beta=18}$},
title style = {align=center, font=\small},
ymajorgrids,
tick label style = {/pgf/number format/fixed},
ymin = 0.39,
ymax = 1.025,
]
\addplot[black,smooth,very thick,loosely dotted] coordinates {
(1,0.43088)
(2,0.681667)
(3,0.825256)
(4,0.905399)
(5,0.949399)
(6,0.973361)
(7,0.986049)
(8,0.992767)
};
\addplot[blue,smooth,very thick,dashdotdotted] coordinates {
(1,0.426139)
(2,0.674749)
(3,0.817535)
(4,0.898444)
(5,0.943746)
(6,0.969033)
(7,0.983215)
(8,0.990994)
};
\addplot[ForestGreen,smooth,very thick,loosely dashed] coordinates {
(1,0.476071)
(2,0.729028)
(3,0.861613)
(4,0.930253)
(5,0.96512)
(6,0.983019)
(7,0.991807)
(8,0.996026)
};
\addplot[red,smooth,very thick] coordinates {
(1,0.455943)
(2,0.709394)
(3,0.847752)
(4,0.921212)
(5,0.960035)
(6,0.979762)
(7,0.99004)
(8,0.99513)
};
\end{axis}
\end{tikzpicture}

\begin{tikzpicture}
\begin{axis}[
name = axis1,
width = 0.5\textwidth, 
height = 0.35\textwidth,
xmin = 1,
xmax = 8,
xtick = {0,1,2,3,4,5,6,7,8}, 
title = {Seeded designs; $\alpha=4$; $\boldsymbol{\beta=36}$},
title style = {align=center, font=\small},
ymajorgrids,
xlabel = Value of $p$,
tick label style = {/pgf/number format/fixed},
ymin = 0.26,
ymax = 1.025,
]
\addplot[black,smooth,very thick,loosely dotted] coordinates {
(1,0.299213)
(2,0.513437)
(3,0.666178)
(4,0.775092)
(5,0.847119)
(6,0.897416)
(7,0.93216)
(8,0.956103)
};
\addplot[blue,smooth,very thick,dashdotdotted] coordinates {
(1,0.296846)
(2,0.511137)
(3,0.664984)
(4,0.774637)
(5,0.845883)
(6,0.895644)
(7,0.930549)
(8,0.954252)
};
\addplot[ForestGreen,smooth,very thick,loosely dashed] coordinates {
(1,0.340083)
(2,0.57048)
(3,0.725362)
(4,0.827466)
(5,0.889633)
(6,0.930511)
(7,0.956866)
(8,0.973567)
};
\addplot[red,smooth,very thick] coordinates {
(1,0.325091)
(2,0.549469)
(3,0.703598)
(4,0.807965)
(5,0.877012)
(6,0.922767)
(7,0.95115)
(8,0.969238)
};
\end{axis}

\begin{axis}[
at = {(axis1.south east)},
xshift = 0.1\textwidth,
width = 0.5\textwidth,
height = 0.35\textwidth,
xmin = 1,
xmax = 8,
xtick = {0,1,2,3,4,5,6,7,8}, 
title = {Unseeded designs; $\alpha=4$; $\boldsymbol{\beta=36}$},
title style = {align=center, font=\small},
ymajorgrids,
xlabel = Value of $p$,
tick label style = {/pgf/number format/fixed},
legend entries = {$KO \qquad$,$G64 \qquad$,$G66 \qquad$,$G46$},
legend style = {at={(-0.2,-0.35)},anchor=north,legend columns = 4,font=\small},
ymin = 0.26,
ymax = 1.025,
]
\addplot[black,smooth,very thick,loosely dotted] coordinates {
(1,0.300422)
(2,0.514582)
(3,0.665566)
(4,0.771021)
(5,0.844355)
(6,0.895189)
(7,0.930072)
(8,0.9537)
};
\addplot[blue,smooth,very thick,dashdotdotted] coordinates {
(1,0.294959)
(2,0.505344)
(3,0.655582)
(4,0.761254)
(5,0.83517)
(6,0.887286)
(7,0.92335)
(8,0.948272)
};
\addplot[ForestGreen,smooth,very thick,loosely dashed] coordinates {
(1,0.332868)
(2,0.557302)
(3,0.708281)
(4,0.809287)
(5,0.87635)
(6,0.920225)
(7,0.948958)
(8,0.967541)
};
\addplot[red,smooth,very thick] coordinates {
(1,0.318502)
(2,0.539071)
(3,0.690688)
(4,0.79442)
(5,0.864153)
(6,0.910994)
(7,0.942052)
(8,0.962845)
};
\end{axis}
\end{tikzpicture}

\end{figure}


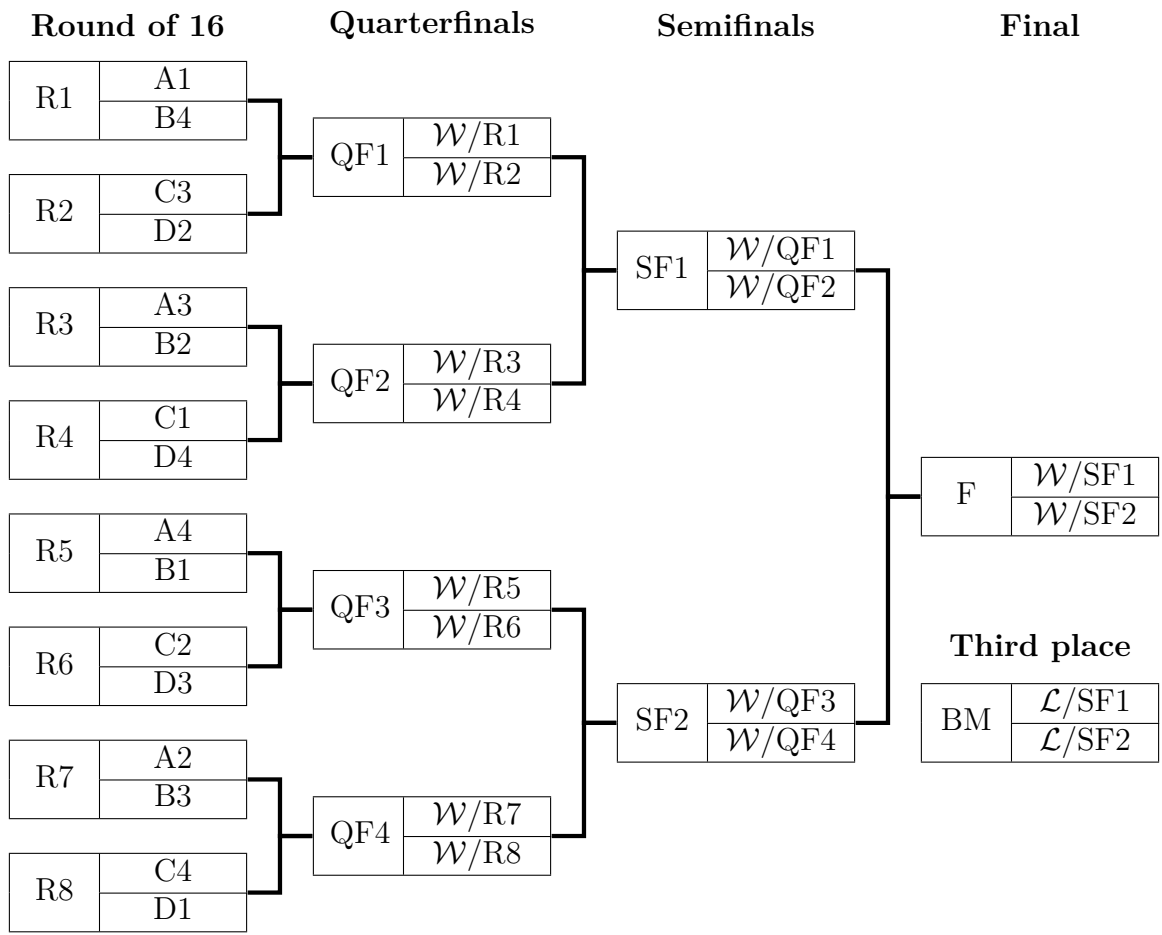
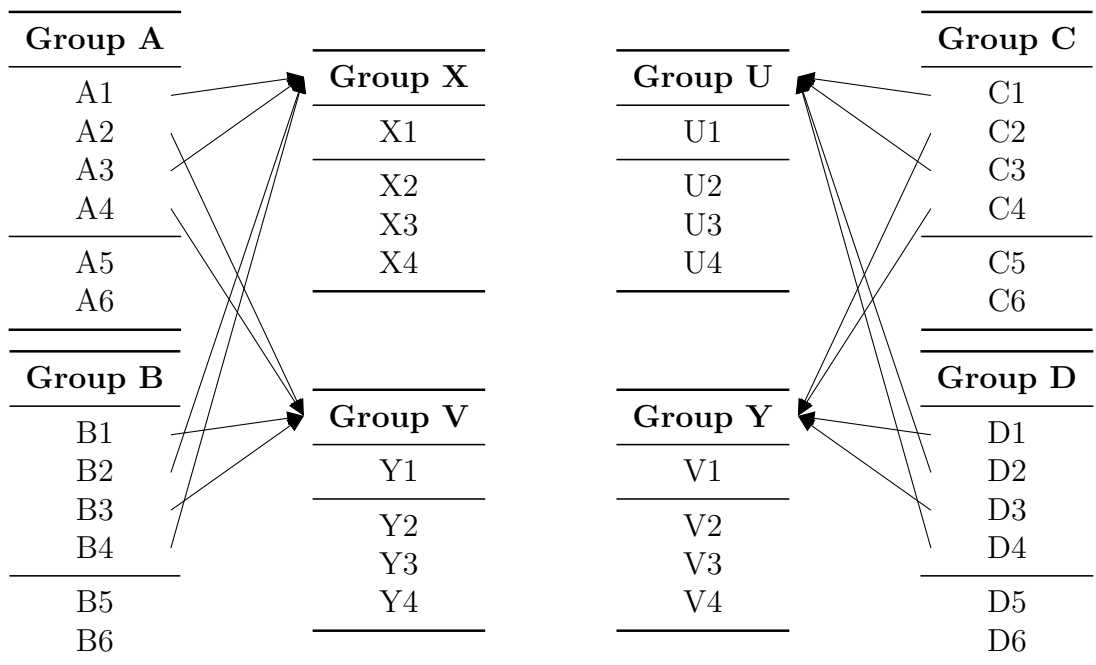
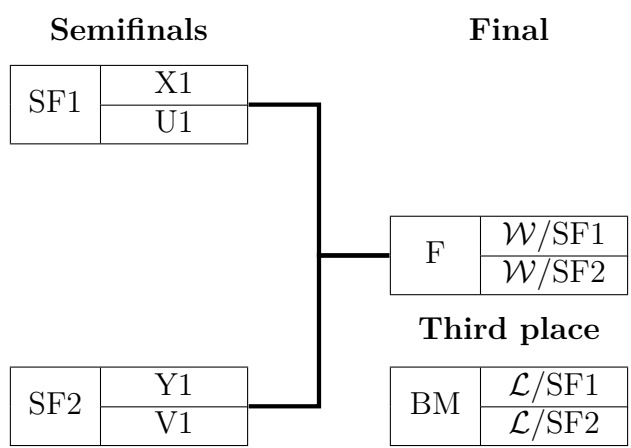
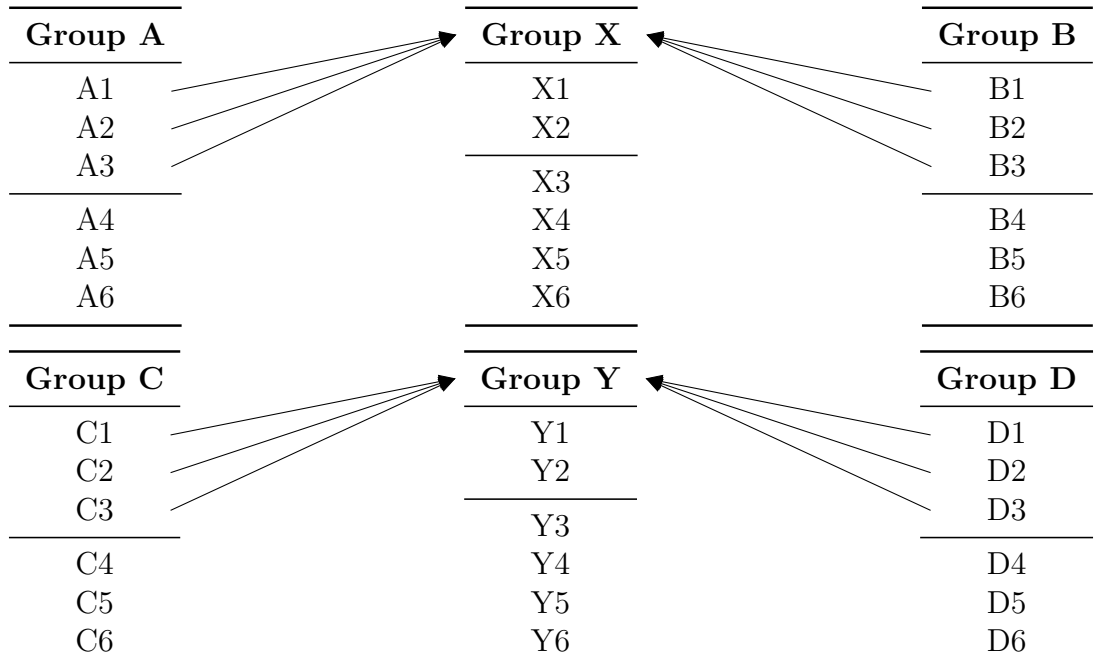
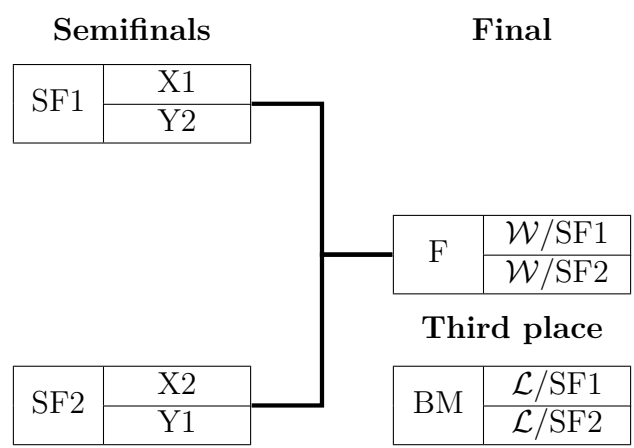
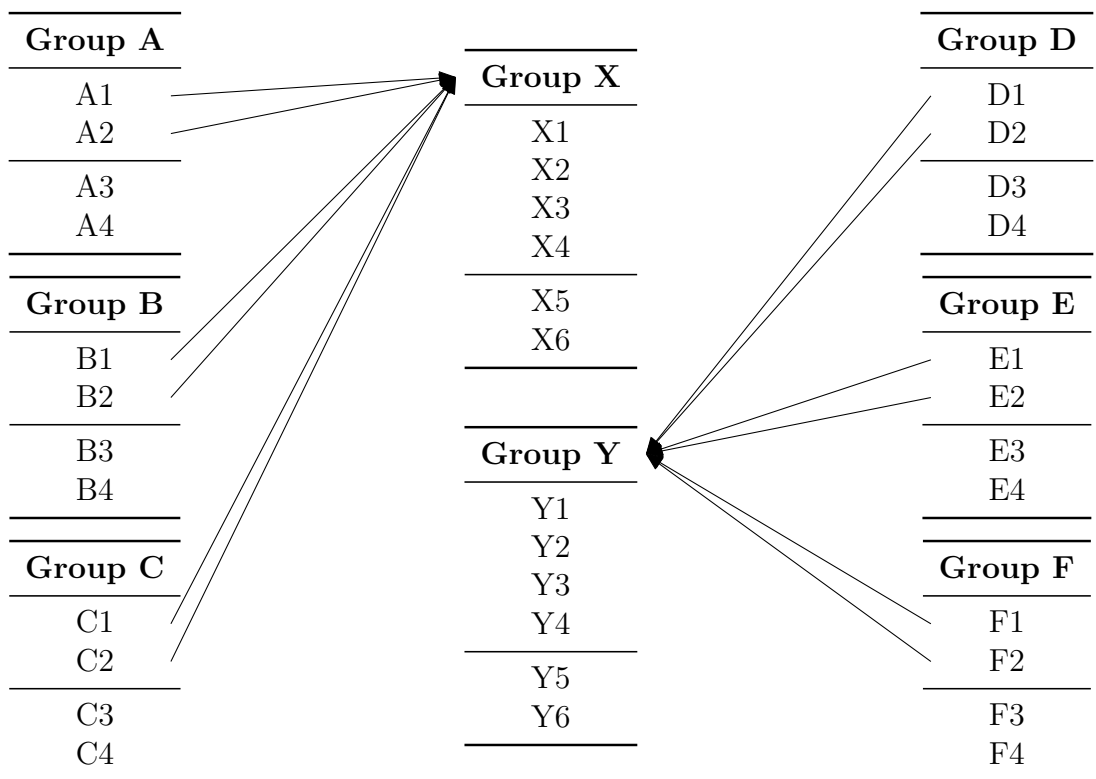
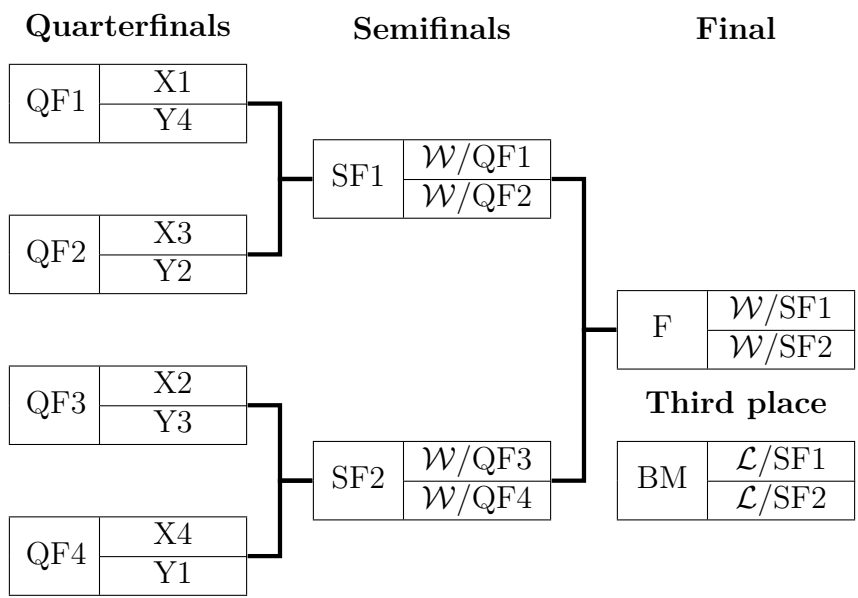
\end{document}